\documentclass[aps,prx,longbibliography,twocolumn,10pt,superscriptaddress,amsmath,amssymb]{revtex4-2}
\usepackage[colorlinks=true, allcolors=blue]{hyperref}
\usepackage{graphicx,dcolumn,bm,nicefrac,times,bbm,braket,enumitem}

\def\im{\mathrm{i}}
\def\dm{\mathrm{d}}
\def\expm{\mathrm{e}}
\def\Hh{\hat{H}}

\def\ch{\hat{c}}
\def\dch{\hat{d}}
\def\nh{\hat{n}}
\def\Ph{\hat{\Psi}}
\def\ph{\hat{\psi}}
\def\pLh{\ph_\textsc{l}}
\def\pRh{\ph_\textsc{r}}
\def\sigmah{\hat{\sigma}}
\def\sigmax{\hat{\sigma}^x}
\def\sigmay{\hat{\sigma}^y}
\def\sigmaz{\hat{\sigma}^z}
\def\sigmap{\hat{\sigma}^{+}}

\def\sigmapm{\hat{\sigma}^{\pm}}
\newcommand{\ssub}[1]{{\scalebox{0.5}{$\mathrm{#1}$}}}

\begin{document}

\title{Theory of topological spin Josephson junctions}
\author{Pei-Xin Shen}
\affiliation{Institute for Interdisciplinary Information Sciences, Tsinghua University, Beijing 100084, China}
\author{Silas Hoffman}
\affiliation{Department of Physics, University of Florida, Gainesville, Florida 32611, USA}
\affiliation{Quantum Theory Project, University of Florida, Gainesville, Florida 32611, USA}
\affiliation{Center for Molecular Magnetic Quantum Materials, University of Florida, Gainesville, Florida 32611, USA}
\author{Mircea Trif}
\email{trifmircea@gmail.com}
\affiliation{Institute for Interdisciplinary Information Sciences, Tsinghua University, Beijing 100084, China}
\affiliation{International Research Centre MagTop, Institute of Physics, Polish Academy of Sciences, Aleja Lotnikow 32/46, PL-02668 Warsaw, Poland}
\date{\today}

\begin{abstract}

We study the spin transport through a 1D quantum Ising-XY-Ising spin link that emulates a topological superconducting-normal-superconducting structure via Jordan-Wigner (JW) transformation. We calculate, both analytically and numerically, the spectrum of spin Andreev bound states and the resulting $\mathbb{Z}_2$ fractional spin Josephson effect (JE) pertaining to the emerging Majorana JW fermions. Deep in the topological regime, we identify an effective time-reversal symmetry that leads to $\mathbb{Z}_4$ fractional spin JE in the \textit{presence} of interactions within the junction. Moreover, we uncover a hidden inversion time-reversal symmetry that protects the $\mathbb{Z}_4$ periodicity in chains with an odd number of spins, even in the \textit{absence} of interactions. We also analyze the entanglement between pairs of spins by evaluating the concurrence in the presence of spin current and highlight the effects of the JW Majorana states. We propose to use a microwave cavity setup for detecting the aforementioned JEs by dispersive readout methods and show that, surprisingly, the $\mathbb{Z}_2$ periodicity is immune to \textit{any} local magnetic perturbations. Our results are relevant for a plethora of spin systems, such as trapped ions, photonic lattices, electron spins in quantum dots, or magnetic impurities on surfaces. 

\end{abstract}
\maketitle

\section{Introduction}

Condensed-matter systems provide an endless playground for emergent exotic phenomena and quasi-particles. In particular, the concept of topological phases associated with the band structure of solids has seen tremendous developments over the past decades \cite{WenRMP2017}. Topological insulators and superconductors are probably among the most scrutinized, notably because they can host Majorana fermions, quasi-particles that are their own antiparticle, which occur as excitations in such materials \cite{HasanRMP2010,QiRMP2011,BeenakkerARCMP2013}. Thanks to non-Abelian statistics, Majorana fermions are crucial ingredients for a functional topological quantum computer: a set of distant, non-interacting Majorana fermions allow, through the process of braiding, to implement a category of topologically protected gates, albeit not universal \cite{NayakRMP2008,AliceaNP2011,LeijnseSST2012}.

Compounds hosting topological superconductivity are rare, for example, $\mathrm{Sr_2RuO_4}$ is believed to be one \cite{MackenzieRMP2003}. However, material engineering of heterostructures composed of semiconducting and superconducting materials can lead to such special superconductors, i.e., 1D nanowires and 2D topological insulators with strong spin-orbit interaction (SOI) proximitized with the conventional $s$-wave superconductor \cite{FuPRL2008,StanescuPRB2010,AliceaRPP2012}. On the other hand, quantum magnets can mimic electronic systems without the proximity requirements \cite{NiuPRB2012,TsvelikPRL2013,GiulianoNPB2016}. Specifically, a 1D quantum Ising model can emulate a Kitaev $p$-wave superconductor, via the renowned Jordan-Wigner transformation (JWT) \cite{JordanZP1928,LiebAP1961,BarouchPRA1970,KitaevP2001}. In particular, the topological phase transition and the occurrence of Majorana fermions as low-energy modes are all mapped into the spin system when the applied transverse magnetic field is varied, where the ferromagnetic (paramagnetic) phase in the spin chain corresponds to the topological (trivial) phase of the fermionic system \cite{Sachdev2011}.

However, one should not be misled: Although there are some analogies of low-energy excitations between fermionic system and spin space, some topological properties will be lost after transformation \cite{TserkovnyakPRA2011,FendleyJSM2012,BackensPRB2017}. In the spin space, Majorana fermions are not localized objects anymore, and they can be mixed simply by a magnetic field along the Ising axis, i.e. the parity of the ground state is fragile. Nevertheless, it is of crucial importance to investigate which of the topological properties associated with Majorana fermions can survive in the spin chain and provide experimental witnesses of their manifestations. To achieve that, in this paper we propose and study the spin transport through an Ising-XY-Ising (IXI) inhomogeneous spin chain in which the Ising axes are misaligned. Borrowing from the electronic description, such a spin chain system emulates a phase-biased topological superconducting-normal-superconducting (SNS) junction that hosts Andreev bound states (ABSs), with a supercurrent flowing through the normal part \cite{Kopnin2001,BeenakkerPRL1991,KwonEPJB2004,Martinis2004}. 

The symmetries of a system play an essential role in the topological phase classification. Nowadays, non-interacting fermionic systems are classified into ten classes by means of three fundamental symmetries: time-reversal symmetry (TRS), particle-hole symmetry (PHS), and sublattice symmetry \cite{KitaevACP2009,WenPRB2012,RyuPRB2012,LudwigPS2015}. In addition, crystalline symmetries (e.g., inversion symmetry) \cite{FuPRB2007,LiuPRL2012,ZhangPRL2013,ZhangPRB2014}, as well as many-body interactions \cite{FidkowskiPRB2010,FidkowskiPRB2011}, can also lead to different topological classes, which, combined with magnetic impurities \cite{PengPRL2016,HuiPRB2017,Vinkler-AvivPRB2017}, may result in various types of Josephson effects (JEs) in superconducting junctions. Roughly speaking, periodicities of JEs are $2\pi$ in the trivial phase, $4\pi$ in the topological phase, and $8\pi$ in the topological phase with many-body interactions or impurities (see Sec.~\ref{Sec:FSJE} for rigorous descriptions). The latter two cases are known as $\mathbb{Z}_2$ and $\mathbb{Z}_4$ fractional JEs pertaining to contributions from Majoranas and parafermions, respectively \cite{Vinkler-AvivPRB2017,ZhangPRL2014,OrthPRB2015}. In this paper, we realize the spin chain-analogs of these JEs and unveil an exotic dependence of the $\mathbb{Z}_2$ and $\mathbb{Z}_4$ fractional spin JEs on the parity of the number of sites. We go on to find several symmetries in the spin chain that protect the associated spin current from various types of spin-spin interactions and demonstrate their robustness against fluctuating magnetic fields.

One of the most counterintuitive characteristics in the quantum world is entanglement whose non-locality provides another instructive insight to understand topological phases \cite{KitaevPRL2006}. Nowadays, there is still no universal way to quantify the entanglement of a mixed state shared by arbitrary subsystems \cite{HorodeckiRMP2009}. However, one can compute the entanglement of a mixed state in a bipartite spin-$\nicefrac{1}{2}$ systems using concurrence \cite{WoottersQIC2001}. The variation of the entanglement across the quantum phase transition point has been investigated in the anisotropic XY spin chain with periodic boundary conditions \cite{OsbornePRA2002}. Here, we evaluate the entanglement between spins and show that it can be enhanced in the presence of a spin current owing to the misaligned Ising axes. This effect, while present in the spin chain, does not have a fermionic counterpart in topological superconductors.

The experimental method of choice for detecting spin current in insulating (quantum) magnets is via the inverse spin-Hall effect in which a metal with strong SOI is coupled to the insulating magnet. Spin current is injected into the metal which is converted, via the SOI, into charge current and can be measured by usual means \cite{SinovaRMP2015}. While this method is effective for large spin systems, the signal might be too small for quantum spin chains. Thus, we propose detecting the spin current by embedding our system in a cavity QED setup wherein such a spin flow shifts both the cavity frequency and the $Q$ factor, which can then be detected by measuring the spectral features of the cavity.

The paper is organized as follows. In Sec.~\ref{Sec:Hamiltonian}, we introduce the spin system and the model Hamiltonian. There we perform the mapping from spins to fermions via the JWT. In Sec.~\ref{Sec:Symmetry}, we analyze the symmetries of the two representations appearing at the lattice level. In Sec.~\ref{Sec:LowEnergy}, we focus on the low-energy sector using both a continuum theory as well as the full lattice diagonalization, to solve for the ABSs spectra analytically, and compare to those found numerically. In Sec.~\ref{Sec:FSJE}, we discuss different scenarios of fractional JEs regarding an effective TRS in the continuum limit and an inversion TRS at the lattice level, respectively. In Sec.~\ref{Sec:Entanglement}, we calculate the texture of the spin entanglement in the presence of a spin supercurrent in the XY sector. In Sec.~\ref{Sec:cQED}, we propose and analyze the coupling of the spin chain to a microwave cavity for readout of the spin current and the periodicities of the JEs, along with examining the robustness of the fractional JEs under perturbations of the in-plane magnetic fields. Finally, in Sec.~\ref{Sec:Conclusions} we conclude with an outlook on future directions.

\section{Model Hamiltonian}
\label{Sec:Hamiltonian}

\begin{figure}
  \centering
  \includegraphics[width=0.48\textwidth]{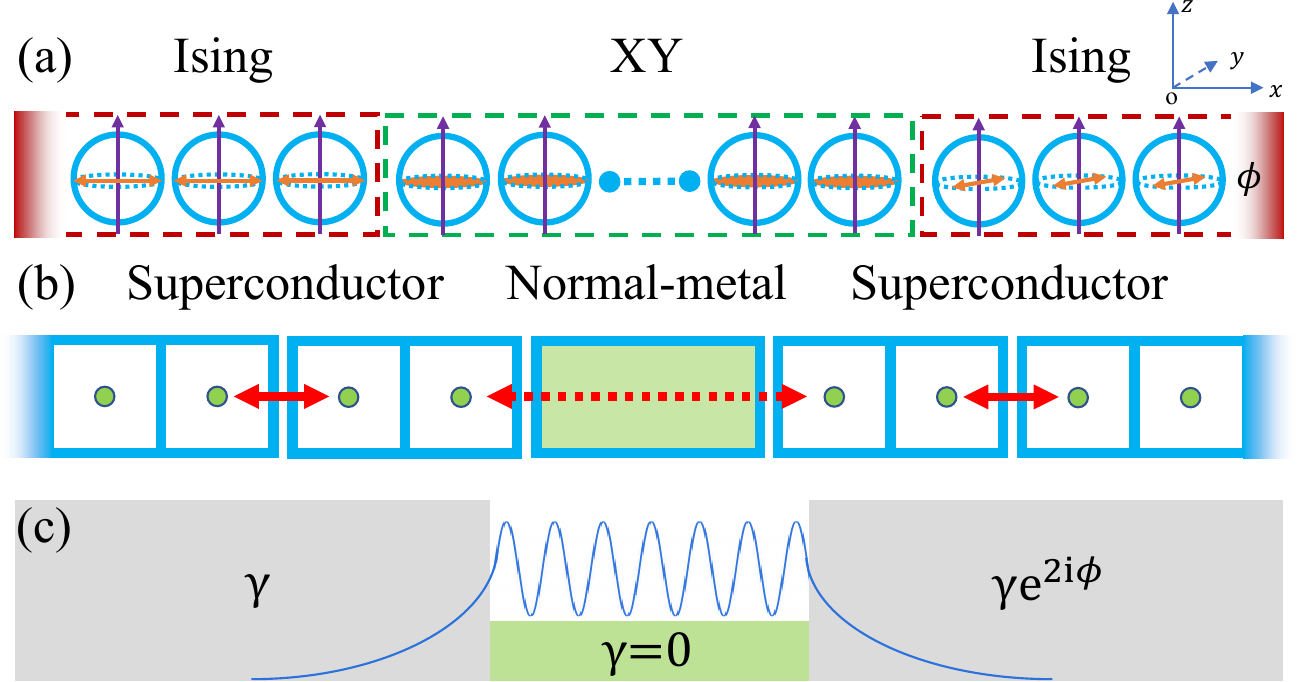}	
  \caption{(a) Schematic of the IXI spin (blue ball) chain in the transverse field (purple arrow): the middle part (green dashed box) is the isotropic XY model, the left and right parts (red dashed boxes) are the quasi-Ising model with the same anisotropy $\gamma$, whereas the right part contains a different spin anisotropic angle $\phi$ (the orientation of the orange arrow) from the left part. (b) After the JWT, the IXI emulates a topological SNS structure, every fermion (blue box) is split into two Majoranas (green dots). There can host Majorana couplings (red dashed arrow) between the left and right $p$-wave superconductors. (c) The wave function (blue curve) of the JW Majorana bound state lies in the gapped-gapless-gapped topological SNS structure.}
  \label{Fig:Schematic}
\end{figure}

The $N$-site anisotropic XY spin chain in a transverse field, presented schematically in Fig.~\ref{Fig:Schematic}(a) with open boundary conditions, is described by the Hamiltonian
\begin{equation} \label{Eqn:HSpinGeneral}
\!\!\Hh_\textsc{g}^\textsc{s}=-J
	\sum\nolimits_i (t_i+\gamma_i) \sigmah^m_i \sigmah^m_{i+1}
	+ (t_i-\gamma_i) \sigmah^{n}_i \sigmah^{n}_{i+1} + g_i\sigmaz_i\,,
\end{equation}
where $\sigmah^{m(n)}_i = \bm{\sigmah}_i \cdot \mathbf{m}_i (\mathbf{n}_i)$, $\bm{\sigmah}_i = (\sigmax_i, \sigmay_i, \sigmaz_i)$ is a spin vector constructed by Pauli matrices at site $i$, $\mathbf{m}_i = (\cos\phi_i, \sin \phi_i, 0)$, $\mathbf{n}_i = (-\sin \phi_i, \cos\phi_i, 0)$, $\phi_i$ is the spin anisotropic angle with respect to the $z$ axis, $0 \leq \gamma_i \leq1$ marks the degree of anisotropy in the $xy$-plane, $J>0$ is the spin exchange constant, $0 \leq t_i \leq1$ is the coupling strength, $g_i=g$ is the relative magnitude of the global transverse field along the $z$ axis. Lengths are measured in units of the lattice spacing $a$. 

By tuning the value of parameters, the chain is split into three regions: Ising-XY-Ising, pertaining to the spin Josephson junctions (JJs). The number of sites in the left, middle, and right parts is $N_\textsc{l}, N_\textsc{m}, N_\textsc{r}$, respectively. The middle chain and the left and right interfaces ($x_\textsc{l}=N_\textsc{l}$, $x_\textsc{r}=N_\textsc{l}+N_\textsc{m}$, respectively) are described by the isotropic XY model by setting $\gamma_i=\phi_i=0$, $\forall i\in [x_\textsc{l},x_\textsc{r}]$. The left and right parts are misaligned, i.e., $\phi_i=\phi$ in the right part and $\phi_i=0$ in the left part, quasi-Ising (anisotropic XY) spin chains such that $\gamma_i=\gamma\neq0$. Although the coupling strength is set as $t_i=t$ in the bulk regions of the spin chain, the parameters at the two interfaces are different: $t_{x_\textsc{l}}=t_{x_\textsc{r}}=\mathbbm{t}$. When $\mathbbm{t}=t$, the connection between different regions are perfect, while if $\mathbbm{t}=0$ they are decoupled from each other.

We perform the JWT, $\ch^\dagger_i = \prod_{j=1}^{i-1} (-\sigmaz_j) \sigmap_i$, $\sigmapm_i = (\sigmax_i \pm \im \sigmay_i)/2$, on Eq.~\eqref{Eqn:HSpinGeneral} and obtain the fermionic Hamiltonian 
\begin{align} \label{Eqn:HFermiGeneral}
  \Hh_\textsc{g}^\textsc{f}=
  &-2J\sum\nolimits_i \big[ (t_i\ch^\dagger_i \ch_{i+1} + \gamma_i \expm^{-2\im\phi_i} \ch^\dagger_{i}\ch^\dagger_{i+1} + \mathrm{H.c.}) \nonumber\\
  &+ g_i (\ch^\dagger_{i} \ch_{i} - 1/2)\big] \,,
\end{align}
where $\ch_i^\dagger$ ($\ch_i$) is the creation (annihilation) operator of the JW electron at site $i$. It turns out the IXI emulates a topological superconducting-normal-superconducting (SNS) junction [Fig.~\ref{Fig:Schematic}(b)]. Since Eq.~\eqref{Eqn:HFermiGeneral} is quadratic, it can be expressed in Bogoliubov-de Gennes (BdG) form $\Hh_\textsc{g}^\textsc{f}=\hat{C}^\dagger \mathcal{H}_\textsc{g}^\textsc{f} \hat{C}/2$ with
\begin{align} \label{Eqn:HBdGFermiGeneral}
	\mathcal{H}_\textsc{g}^\textsc{f}=
	&-2J\sum\nolimits_i \big\{ [ (t_i\rho_z + \im \gamma_i \expm^{-2\im \phi_i \rho_z} \rho_y ) \otimes \ket{i}\bra{i+1} \nonumber\\
	&+\mathrm{H.c.} ] + g_i\rho_z \otimes \ket{i}\bra{i} \big\} \,,
\end{align}
where $\hat{C} = (\ch_1, \ch_2, \dots, \ch_\textsc{n}, \ch^\dagger_1, \ch^\dagger_2, \dots, \ch^\dagger_\textsc{n})^\textsc{t}$ is a 2$N$-dimensional spinor and $\ket{i} = (0,\dots,1,0,\dots)^\textsc{t}$ is an $N$-dimensional basis vector corresponding to the $i$th site of the chain, and $\rho_{y}$ and $\rho_{z}$ are Pauli matrices acting on the Nambu particle-hole space. By use of the Bogoliubov quasi-particle $\hat{D} = (\dch_1, \dch_2, \dots, \dch_\textsc{n}, \dch^\dagger_1, \dch^\dagger_2, \dots, \dch^\dagger_\textsc{n})^\textsc{t}$ basis, $\Hh_\textsc{g}^\textsc{f}$ is diagonalized into $\sum_n \epsilon_n ( \dch^\dagger_n \dch_n - \nicefrac{1}{2} )$ with a set of single-particle energy $\epsilon_n$. 

When the twisting angle $\phi$ of the right Ising part is nonzero, there is a spin supercurrent flowing through the middle sector, whose coupling Hamiltonian is XY type, $\hat{H}_\textsc{xy}=-Jt(\sigmax_i\sigmax_{i+1} + \sigmay_i\sigmay_{i+1})$. Hence, via evaluating the Heisenberg equations of motion $\Delta\hat{J}_z = \hat{J}^\mathrm{out}_z - \hat{J}^\mathrm{in}_z = \im[\sigmaz_i, \hat{H}_\textsc{xy}]$, we define a $z$-component spin current operator as $\hat{J}_z \equiv \hat{J}^\mathrm{out}_z$ \cite{ChenPRB2013,ZhengPRB2017} or, more explicitly \footnote{Alternatively, one can define $\hat{J}_z \equiv \hat{J}^\mathrm{in}_z = 2Jt(\sigmay_{i-1}\sigmax_{i} - \sigmax_{i-1}\sigmay_{i}) = 4\im Jt(\ch^\dagger_{i}\ch_{i-1} - \ch^\dagger_{i-1}\ch_{i})$, whose expectation value is the same as $\hat{J}^\mathrm{out}_z$, since $\Delta\hat{J}_z = 0$ for a stationary state. },
\begin{equation} \label{Eqn:SpinCurrent}
\hat{J}_z/(-2Jt)= \sigmax_i\sigmay_{i+1} - \sigmay_i\sigmax_{i+1} = 2\im(\ch^\dagger_i\ch_{i+1} - \ch^\dagger_{i+1}\ch_i) \,.
\end{equation}
In this paper, we only focus on $\hat{J}_z$, since the expectation values of $\hat{J}_x$ and $\hat{J}_y$ vanish, while $\hat{J}_z$ remains a constant $\forall i\in (x_\textsc{l},x_\textsc{r})$ in the middle part due to current conservation. Such spin superfluidity is analogous to the superconductivity in the presence of a phase bias: As charge conservation is broken at the level of mean-field theory for superconductors, $\hat{J}_z$ is not conserved in the Ising portions. Thus, the lattice, whose dynamics are neglected in this paper, effectively acts as a source and drain of spin.

\section{Lattice Symmetry Analysis}
\label{Sec:Symmetry}

The symmetries of a system are independent of representations, although they can be interpreted differently in the spin and fermionic pictures. In the following subsections, we will identify the symmetries occurring in the spin system and find out their fermionic counterparts through the JWT. To be more general, we introduce two types of interacting Hamiltonians: the spin-spin interactions in the $z$ direction (ZZ type),
\begin{equation} \label{Eqn:HSpinZZ}
\Hh_\textsc{i}^\textsc{s}= - J \sum\nolimits_i \delta_i \sigmaz_i \sigmaz_{i+1} \,,
\end{equation}
acting on the spin space, and the Coulomb interactions (NN type),
\begin{equation} \label{Eqn:HFermiNN}
\Hh_\textsc{i}^\textsc{f}= - 4 J \sum\nolimits_i \chi_i \nh_i \nh_{i+1} \,.
\end{equation}
Equations~\eqref{Eqn:HSpinZZ} and \eqref{Eqn:HFermiNN} are connected by the JWT up to global renormalization of the magnetic field $4\nh_i \nh_{i+1} \Leftrightarrow 1+\sigmaz_i + \sigmaz_{i+1} + \sigmaz_i \sigmaz_{i+1}$ which, as  we see below, will have significant implications.

\subsection{Spin \texorpdfstring{$\mathbb{Z}_2$}{Z2} Symmetry}

The spin chain has a $\mathbb{Z}_2$ symmetry since $[\Hh_\textsc{g}^\textsc{s}, \hat{P}_\textsc{s}]=0$ with $\hat{P}_\textsc{s} = \prod\nolimits_i \sigmaz_i$, $\hat{P}_\textsc{s}^2=+1$, which acts on Pauli operators as 
\begin{equation}
\hat{P}_\textsc{s} \sigmah^{m(n)}_i \hat{P}_\textsc{s}^{-1} = -\sigmah^{m(n)}_i \,,\quad
\hat{P}_\textsc{s} \sigmaz_i \hat{P}_\textsc{s}^{-1} = +\sigmaz_i \,.
\end{equation}
By the JWT, the corresponding operator in the fermionic system is identified as a parity operator $\hat{P}_\textsc{f} = \prod\nolimits_i (2\nh_i-1)$, which transforms fermionic operators as
\begin{equation}
\hat{P}_\textsc{f} \ch^\dagger_i \hat{P}_\textsc{f}^{-1} = -\ch^\dagger_i \,,\quad
\hat{P}_\textsc{f} \ch_i \hat{P}_\textsc{f}^{-1} = -\ch_i \,.
\end{equation}
Since Eq.~\eqref{Eqn:HFermiGeneral} is a sum of terms containing an even number of fermionic creation and annihilation operators, the system is required to preserve the parity as $[\Hh_\textsc{g}^\textsc{f}, \hat{P}_\textsc{f}]=0$ at any time, although the number of fermions is not conserved. One can easily verify that $\mathbb{Z}_2$ symmetry holds for the aforementioned two types of interacting Hamiltonians in Eqs.~\eqref{Eqn:HSpinZZ}-\eqref{Eqn:HFermiNN}. Considering the pure Ising chain with $\delta_i=g_i=0$ and $\gamma_i=t_i$, the spin ground states will simultaneously break above $\mathbb{Z}_2$ symmetry, which in turn gives two degenerate ground states in the Kitaev model characterized by Majorana zero modes.

\subsection{Real Time-Reversal Symmetry}

If $g_i=0$ globally, Eq.~\eqref{Eqn:HSpinGeneral} contains \textit{real} TRS (rTRS) with $[\Hh_\textsc{g}^\textsc{s}, \hat{T}_\textsc{s}]=0$ by the operator $\hat{T}_\textsc{s}=\prod_i \im \sigmay_i \mathcal{K}$ acting on the Pauli operators as 
\begin{equation}
\hat{T}_\textsc{s} \hat{\sigma}^{\alpha}_i \hat{T}_\textsc{s}^{-1}= -\hat{\sigma}^{\alpha}_i\,, \quad \alpha=m,n,z\,.
\end{equation}
where $\mathcal{K}$ is an anti-unitary complex conjugate operator. Since $\hat{T}_\textsc{s}^2=(-1)^N$, according to Kramers theorem, all many-body spectra must be at least doubly degenerate when $N$ is odd. Through the JWT, Eq.~\eqref{Eqn:HFermiGeneral} fulfills $[\Hh_\textsc{g}^\textsc{f}, \hat{T}_\textsc{f}]=0$ inherited from the spin space, $\hat{T}_\textsc{f} = \prod_{i} [\ch^\dagger_i+(-1)^{N+1+i}\ch_{i}] \mathcal{K}$ is a \textit{second-quantized} operator acting on Fock space as $\hat{T}_\textsc{f} \im \hat{T}^{-1}_\textsc{f} = -\im$ ,
\begin{equation} \label{Eqn:RealTRS2nd}
\hat{T}_\textsc{f} \ch^\dagger_{i} \hat{T}^{-1}_\textsc{f} = (-1)^i \ch_{i}\,, \quad
\hat{T}_\textsc{f} \ch_{i} \hat{T}^{-1}_\textsc{f} = (-1)^i \ch^\dagger_{i}\,.
\end{equation}
This can be interpreted as the charge conjugation in the fermionic language. Based on non-interacting Eq.~\eqref{Eqn:HBdGFermiGeneral}, we can rewrite $\hat{T}_\textsc{f}$ in a \textit{first-quantized} form
\begin{equation} \label{Eqn:RealTRS1st}
\mathcal{T}_\textsc{f} = \rho_x \mathcal{K} \otimes \sum\nolimits_i (-1)^i \ket{i}\bra{i}\,, \quad \mathcal{T}^2_\textsc{f}=+1\,,
\end{equation}
which renders $[\mathcal{H}_\textsc{g}^\textsc{f}, \mathcal{T}_\textsc{f}]=0$. Note that Eqs.~\eqref{Eqn:RealTRS2nd} are more general than Eqs.~\eqref{Eqn:RealTRS1st} since they can accommodate interactions, i.e., Eq.~\eqref{Eqn:HFermiNN}. We find $[\Hh_\textsc{i}^\textsc{f}, \hat{T}_\textsc{f}]\neq0$, yet the ZZ-type interactions in Eq.~\eqref{Eqn:HSpinZZ} retain rTRS due to $[\Hh_\textsc{i}^\textsc{s}, \hat{T}_\textsc{s}]=0$. 

When $N$ is odd, the twofold degeneracies in the many-body spectrum are protected by the second-quantized rTRS operator with $\hat{T}^2_\textsc{f}=-1$, which enforces intrinsic zero modes in the single-particle spectrum. Under the fermionic picture, as the coupling strength $\mathbbm{t}$ increases, amplitudes of the intrinsic zero modes in the middle part will exponentially leak into the superconducting parts, and fully merge with Majorana zero modes in the thermodynamic limit, whose wave functions are well localized at the edges of the chain and cause no effect on the in-gap spectrum.

\subsection{Inversion Time-Reversal Symmetry}

Although the first-quantized rTRS operator $\mathcal{T}^2_\textsc{f}=+1$ cannot reflect any degenerate properties in the single-particle spectrum, it gives us a hint to find out a hidden \textit{inversion} TRS (iTRS) which leads to an odd-even effect (see discussions in Sec.~\ref{Sec:OddEven}). We first introduce a lattice inversion operator, 
\begin{equation}\label{Eqn:InversionTRS1st}
\mathcal{I}=\sum\nolimits_i (-1)^i \left(\ket{i}\bra{N+1-i}\right)\,, \quad \mathcal{I}^2=(-1)^{N+1}\,,
\end{equation}
which will transform matrix elements of the nearest-neighboring sites with an additional minus sign after applying on the lattice space, e.g., $\tilde{t}_i \equiv t_{N-i} \rightarrow -t_i $ whereas $\tilde{g}_i \equiv g_{N+1-i} \rightarrow g_i$, where we denote parameters with tilde are elements inverted from original position. With the help of $\mathcal{I}$, we can define the iTRS operator
\begin{equation}
\mathcal{T}_\textsc{i}=
\begin{cases}
\im \rho_y \mathcal{K} \otimes \mathcal{I}\,, &\text{for odd $N$}\\
\rho_x \mathcal{K} \otimes \mathcal{I}\,, &\text{for even $N$ \,.}
\end{cases}
\end{equation}
Applying $\mathcal{T}_\textsc{i}$ to Eq.~\eqref{Eqn:HBdGFermiGeneral} as $\mathcal{T}_\textsc{i}\mathcal{H}_\textsc{g}^\textsc{f}\mathcal{T}_\textsc{i}^{-1}$, we obtain
\begin{align}\label{Eqn:HBdGFermiGeneralTRS}
	-2J\sum\nolimits_{i}
	& \big\{ \big[ \big( \tilde{t}_i\rho_z - (-1)^N\; \im \tilde{\gamma}_i \expm^{ - 2\im \tilde{\phi}_i \rho_z}\rho_y \big)\nonumber\\
	& \otimes \ket{i}\bra{i+1} + \mathrm{H.c.} \big]- \tilde{g}_i\rho_z \otimes \ket{i}\bra{i} \big\}\,.
\end{align}
By comparing Eqs.~\eqref{Eqn:HBdGFermiGeneral} and \eqref{Eqn:HBdGFermiGeneralTRS}, it turns out that to retain iTRS as $[\mathcal{H}_\textsc{g}^\textsc{f} (\phi), \mathcal{T}_\textsc{i}]=0$ in the IXI chain, not only should we set $\tilde{g}_i=g_i=0$, $\tilde{t}_i=t_i$, $\tilde{\gamma}_i=\gamma_i$, but also $\phi$ is restricted to the following values:
\begin{equation}\label{Eqn:InversionTRSphi}
\mkern-8mu \expm^{-2\im\phi}=(-1)^{N+1} \Leftrightarrow \phi=
\begin{cases}
\quad\; l\pi \,, &\text{for odd $N$}\\
\pi/2 + l\pi \,, &\text{for even $N$\,,}
\end{cases}
\end{equation}
with $l\in \mathbb{Z}$. Note that $\mathcal{T}_\textsc{i}^2=-1$ in both odd-even cases, according to Kramers theorem, all \textit{single-particle} states at above specific $\phi$ should contain twofold degeneracy.

More generally, we rewrite Eqs.~\eqref{Eqn:InversionTRS1st} in a second-quantized form acting on fermions as $\hat{T}_\textsc{i} \im \hat{T}^{-1}_\textsc{i} = -\im$ ,
\begin{equation} \label{Eqn:InversionTRS2nd}
\hat{T}_\textsc{i} \ch^\dagger_{i} \hat{T}^{-1}_\textsc{i} =
\begin{cases}
 +\im(-1)^i \ch_{N+1-i}\,, & \text{for odd $N$}\\
(-1)^i \ch_{N+1-i}\,, & \text{for even $N$\,,}
\end{cases}
\end{equation}
and its actions on spins are $\hat{T}_\textsc{i} \sigmaz_i \hat{T}_\textsc{i}^{-1} = -\sigmaz_{N+1-i}$,
\begin{equation} \label{Eqn:InversionTRSspin}
\hat{T}_\textsc{i} \sigmah^{m(n)}_i \hat{T}_\textsc{i}^{-1} = 
\begin{cases}
+\im\hat{P}_\textsc{s}\sigmah^{m(n)}_{N+1-i}\,, & \text{for odd $N$}\\
\pm\im\hat{P}_\textsc{s}\sigmah^{n(m)}_{N+1-i}\,, & \text{for even $N$\,,}
\end{cases}
\end{equation}
which can be understood as the charge-parity symmetry. Applying the rTRS operator to the ZZ-type interactions of Eq.~\eqref{Eqn:HSpinZZ}, we get
\begin{equation}
\hat{T}_\textsc{i} (-\Hh_\textsc{i}^\textsc{s}/J) \hat{T}^{-1}_\textsc{i}=
\sum\nolimits_i \tilde{\delta}_{i} \sigmaz_{i} \sigmaz_{i+1}\,.
\end{equation}
Once $\delta_i=\tilde{\delta}_i$ are set symmetrically, the system Hamiltonian always commutes with iTRS operator at specific $\phi$ illustrated in Eq.~\eqref{Eqn:InversionTRSphi}, which ensures twofold degeneracies of many-body states in the interacting case. As for the NN-type interactions in Eq.~\eqref{Eqn:HFermiNN}, we obtain
\begin{equation}
\hat{T}_\textsc{i} (-\Hh_\textsc{i}^\textsc{f}/4J) \hat{T}^{-1}_\textsc{i}=
\sum\nolimits_i \tilde{\chi}_i \ch_{i}\ch^\dagger_{i} \ch_{i+1}\ch^\dagger_{i+1}\,, 
\end{equation}
and expand to $\sum_i \tilde{\chi}_i (\nh_i \nh_{i+1}+1-\ch^\dagger_{i}\ch_{i}-\ch^\dagger_{i+1}\ch_{i+1})$, whose last three terms will break iTRS at any $\phi$, even if we set $\chi_i=\tilde{\chi}_i$ symmetrically. Such seemingly trivial local terms will dramatically alter the periodicities of the spin JEs (see Fig.~\ref{Fig:ContiZZNN} and further discussions in Sec.~\ref{Sec:FSJE}).

Note that in the above proof all parameters are required to hold strict inversion symmetry under $N_\textsc{l}=N_\textsc{r}$, thus the odd-even effect only depends on $N_\textsc{m}$. However, by the fact that the ABSs decay exponentially in the two superconducting parts, as long as their lengths are much larger than superconducting coherence length, the degenerate properties are still robust within the energy gap regardless of the parity and the equality of $N_\textsc{l}$ and $N_\textsc{r}$, which in turn underscores the dominance of $N_\textsc{m}$.

\section{Low-Energy Theory}
\label{Sec:LowEnergy}

In the following subsections, we will focus on the low-energy sectors, with the aid of fermionic descriptions, utilizing both a continuum theory and full lattice diagonalization. Given translation symmetry under periodic boundary conditions, the bulk spectrum of the isolated anisotropic XY spin chain reads \cite{LiebAP1961}
\begin{equation}\label{Eqn:KSpetrum}
\epsilon_k = 2J\sqrt{(2t\cos ka + g)^2 + 4\gamma^2\sin ^2 ka}\,, 
\end{equation}
where $k$ is the wave number after the Fourier transformation. When $\gamma\neq0$, the spectrum is always gapped except at $|g|=2t$ where the system undergoes a quantum phase transition. In the case of $|g|<2t$, the fermionic chain will be in a topological phase where Majorana fermions appear at the edges if we cut off the chain, and the corresponding topological invariant is characterized by the topological winding number $\mathcal{W}=1$ (see Appendix~\ref{Appx:SpinChain} for details). However, if $|g|>2t$ such edge modes will disappear, the chain enters the trivial phase, and the value of the topological winding number goes to zero. Figure~\ref{Fig:Schematic}(c) depicts the wave function of the JW Majorana bound state (MBS) in the presence of a phase bias between two superconducting parts. Note that the middle sector is gapped in the trivial regime $|g|>2t$, which hinders the occurrence of the supercurrent and makes the chain insulating. Since we are interested in the JEs pertaining to the supercurrent, we will only focus on the topological regime in the whole paper.

\subsection{Near the Critical Point}

On account of the long wavelength excitations dominating the low-energy properties near the critical point \footnote{Strictly speaking, the energy gap occurs at $k = 0$ when $g\rightarrow-2t$, while if $g\rightarrow+2t$, the energy gap takes at $k = \pm \pi/a$. The sign of $g$ only depends on the direction of the $z$ axis in the spin Hamiltonian and does not cause any different observational effect. When $g>0$, we can define $k'=k+\pi/a$ to translate the momentum in the Brillouin zone and come back to the case of $g<0$, hence in the whole paper we only investigate the negative regime.}, we can replace the fermionic operators in Eq.~\eqref{Eqn:HFermiGeneral} by a continuous Fermi field operator $\ch_i = \sqrt{a} \ph(x)$ and expand it to second order in the spatial gradients to obtain the single-particle continuous Hamiltonian,
\begin{equation}\label{Eqn:HCriticalGeneral}
	\mkern-8mu \mathcal{H}_\textsc{g}^\textsc{c}/2J=
	-\left(2t+g+t a^2\partial^2_x\right)\rho_z - 2\im \gamma_i a \expm^{-2\im \phi_i \rho_z} \rho_y \partial_x \,,
\end{equation}
where $\mathcal{H}_\textsc{g}^\textsc{c}$ is a matrix in the BdG form, and $\Hh_\textsc{g}^\textsc{c} = \nicefrac{1}{2} \int \dm x \Ph(x)^\dagger \mathcal{H}_\textsc{g}^\textsc{c} \Ph(x)$ with a field spinor $\Ph(x) =[\ph(x),\ph^\dagger(x)]^\textsc{t}$. The coefficient in front of the second and first derivative indicates the effective mass, $m^*_i=\hbar^2/(4Jt a^2)$, and velocity, respectively \cite{Sachdev2011}. To mimic the imperfect connections between different parts, we introduce a fictitious potential $\lambda a \delta(x-x_\textsc{l,r})$ at two interfaces, $x=x_\textsc{l,r}$, with barrier strength $\lambda$. When $\lambda\rightarrow\infty$, the three parts of the chain are decoupled from each other. Through the S-matrix approach \cite{Kopnin2001,Hoffman2018}, we obtain the solvability equation for the ABSs spectrum,
\begin{equation}\label{Eqn:CriticalDET}
  \mathrm{Re}\left[\mathcal{S}_0^2 \expm^{\im (\mathrm{K}_\textsc{m}^\textsc{+}-\mathrm{K}_\textsc{m}^\textsc{-}) L} - \mathcal{S}_1^2 \expm^{\im (\mathrm{K}_\textsc{m}^\textsc{+}+\mathrm{K}_\textsc{m}^\textsc{-}) L}\right]=\mathcal{S}_2^2 \cos(2\phi)\,,
  \end{equation}
where $\mathrm{K}_\textsc{m}^\ssub{\pm}=\sqrt{\Omega\pm\Xi}/ta$ are the middle wave numbers with $\Omega=t(2t+g)$, $\Xi=t\epsilon/2J$, and $L=(N_\textsc{m}+1)a$ is the length of the middle part, $\mathcal{S}_{0,1,2}$ are the entries of the S matrix, whose explicit expressions are given in Appendix~\ref{Appx:CriticalWFs}, together with the wave functions and the technical details. In the leading order series expansion around zero energy, the spectrum $\mathrm{E}=\Xi/t$ is given by
\begin{equation}\label{Eqn:CriticalLeading}
  \mkern-10mu
  \mathrm{E}=2\sqrt{\Omega}\left(\frac{\pi}{2}\mp\phi+n\pi\right)\Big/\left[\frac{L}{a}+\frac{t(\lambda-\gamma)^2+2t\Omega}{2\gamma\Omega}\right],
\end{equation}
which is plotted in Fig.~\ref{Fig:Spectrum}(a) against the spectra from the exact continuum theory and the lattice model.

\subsection{Deep Topological Regime}

\begin{figure}
  \centering
  \includegraphics[width=0.48\textwidth]{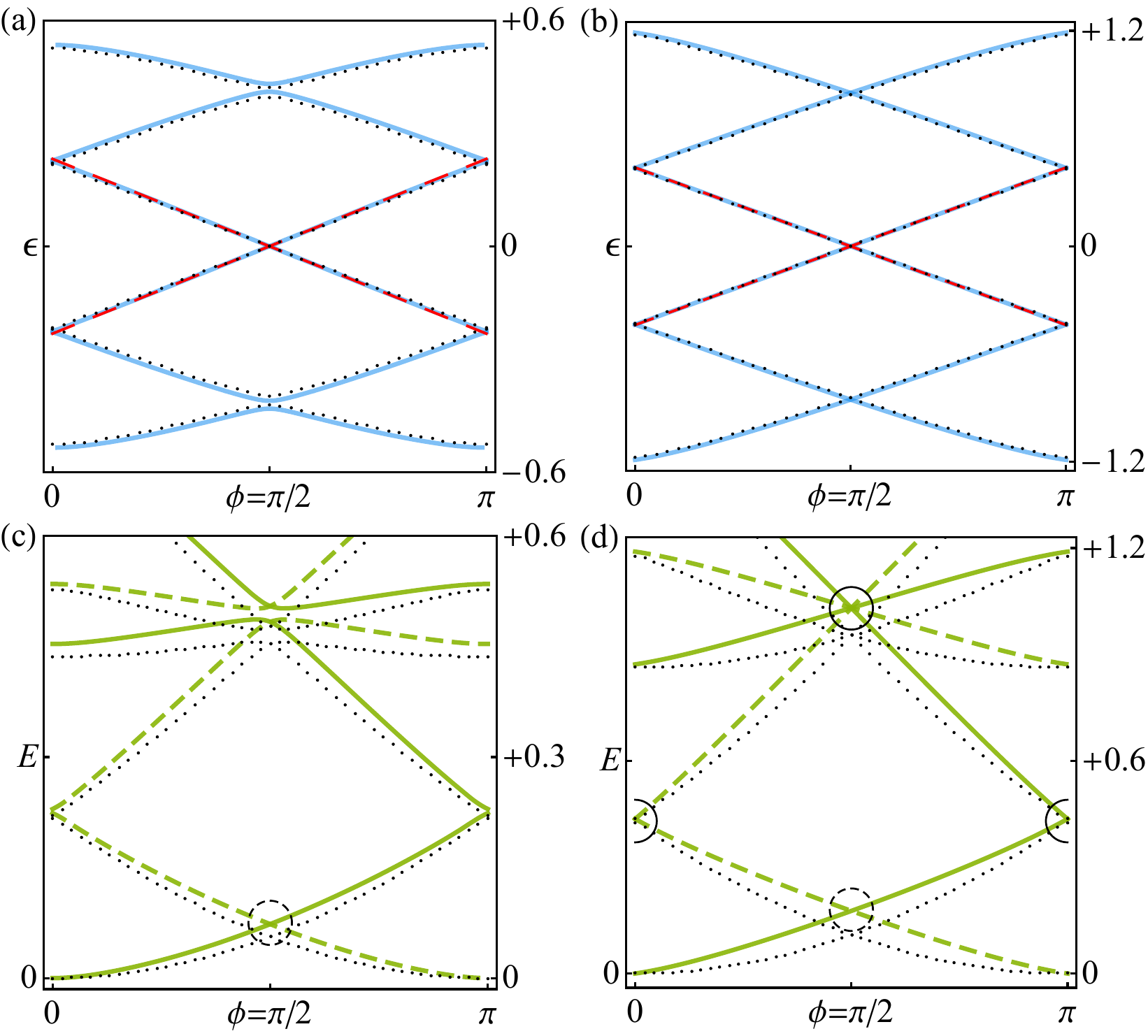}
  \caption{Spectra of the IXI as a function of $\phi$ by evaluating $J=a=t=\mathbbm{t}=1$, $\gamma=0.3$, $\lambda=0$, $N_\textsc{l}=N_\textsc{r}=100$, $N_\textsc{m}=10$ in two regimes. (a) and (b) are the single-particle spectra near the critical point $g=-1.7$ and in the deep topological regime $g=0$, respectively. (c) and (d) are their corresponding many-body spectra. Black dots are solved numerically by the lattice model, blue lines are obtained by solving the transcendental Eqs.~\eqref{Eqn:CriticalDET} and \eqref{Eqn:DeepDET} of the low-energy continuum theory in two regimes, red dashed lines are Majorana solutions calculated by the explicit Eqs.~\eqref{Eqn:CriticalLeading} and \eqref{Eqn:DeepLeading} by setting $n=-1$ $(n=0)$ for upward (downward) branch, green (dashed) lines are many-body spectra with even (odd) parity constructed by the single-particle energies, solid (dashed) circles are crossings protected by the eTRS (PHS).}
  \label{Fig:Spectrum}
\end{figure}

In the deep topological regime  $g\rightarrow0$, the energy gap $\epsilon_\ssub{gap} = 2J \gamma \sqrt{4-g^{2}/(t^2-\gamma^{2})} \rightarrow 4J\gamma$ occurs around $\pm k_\textsc{f} = \pm \arccos (-g/2t )/a \approx \pm \pi/2a$ with the proviso of $\gamma \ll t$. Accordingly, we can expand the lattice fermionic operator around two Fermi points as $\ch_{i}/\sqrt{a} = \expm^{+\im k_\textsc{f} x} \pRh(x) + \expm^{-\im k_\textsc{f} x} \pLh(x)$, where $\ph_\ssub{R,L}$ are right and left mover field operators. We substitute the above transformation into Eq.~\eqref{Eqn:HFermiGeneral}, expand it to the leading order in the spatial gradients and neglect the fast oscillating terms. By defining a continuous Fermi field spinor $\Ph(x) =[\pRh(x),\pLh(x),\pLh^\dagger(x),-\pRh^\dagger(x)]^\textsc{t}$, the deep topological Hamiltonian can be expressed in the BdG form $\Hh_\textsc{g}^\textsc{d} = \nicefrac{1}{2} \int \dm x \Ph(x)^\dagger \mathcal{H}_\textsc{g}^\textsc{d} \Ph(x)$ with matrix $\mathcal{H}_\textsc{g}^\textsc{d}$ as
\begin{equation}\label{Eqn:HDeepGeneral}
	\mathcal{H}_\textsc{g}^\textsc{d}/2J=
	\Upsilon(-\im \partial_x) \rho_z \tau_z + \Delta_i \expm^{ - 2\im \phi_i \rho_z} \rho_x\,,
\end{equation}
where $\Upsilon = 2t a\sin(k_\textsc{f} a)$ is the effective velocity, $\Delta_i = 2\gamma_i\sin(k_\textsc{f} a)$ is the effective pairing potential \cite{Kopnin2001}, $\tau_{x,y,z}$ are Pauli matrices acting on the mover space. Note that the phase is globally shifted by $\pi/4$ in order to keep $\Delta_i$ a real number. The above Hamiltonian shares the same form with JJs created at the edge of a quantum spin Hall (QSH) insulator \cite{BernevigS2006,KonigS2007,FuPRB2009}: our movers $\ph_\ssub{R,L}$ in $\vec{\tau}$ space correspond to their two edge states living in the spin space. Hence, the IXI chain emulates the QSH JJs at low energies. The rTRS in the QSH JJ equates to an \textit{effective} TRS (eTRS) in the IXI chain with $[\mathcal{H}_\textsc{g}^\textsc{d} (\phi)$, $\mathcal{T}_\textsc{e}] = 0$, $\mathcal{T}_\textsc{e}=\im\tau_y\mathcal{K}$ at $\phi=l\pi/2, l\in \mathbb{Z}$ \cite{LopesPRB2019}. Since $\mathcal{T}_\textsc{e}^2=-1$, there must be spectrum degeneracies at those specific phases due to Kramers theorem. With the help of the S-matrix technique, we obtain the transcendental equation for the ABSs in the deep topological regime:
\begin{equation}\label{Eqn:DeepDET}
\mathrm{E}\,L/\Upsilon +\tau \phi=\arccos(\mathrm{E}/\Delta)+n\pi,\quad n \in \mathbb{Z}\,.
\end{equation}
Under the low-energy leading approximation, the energy can be expressed explicitly as
\begin{equation}\label{Eqn:DeepLeading}
\mathrm{E} = (\pi/2-\tau\phi+n\pi)/(L/\Upsilon+1/\Delta),\quad n \in \mathbb{Z}\,,
\end{equation}
which is plotted in Fig.~\ref{Fig:Spectrum}(b) against the exact continuum spectrum and the full lattice spectrum. The index $\tau$ indicates the slope of the spectrum as a function of $\phi$: $\tau=\pm1$ for the downward (upward) branches respectively. In the case of the point contact limit $L\rightarrow0$, Eq.~\eqref{Eqn:DeepDET} is reduced to $\mathrm{E} \rightarrow \tau\Delta\cos\phi$ \cite{KwonEPJB2004}. In Appendix~\ref{Appx:DeepWFs}, we present the explicit wave functions and the technical details of the S matrix.

\subsection{Lattice Diagonalization}

With the single-particle spectrum $\epsilon_n$ solved exactly from the numerical lattice diagonalization, we can construct the many-body spectrum $E_n$: The ground state is built with all the negative-energy single-particles filled, the following excited states are obtained by adding the corresponding quasi-particles to the ground state, whose total number characterizes the parity of the system. Note, however, we could only utilize a few ABSs to create the many-body spectra from the continuum theory.

Figure~\ref{Fig:Spectrum} displays the exact numerical single-particle and the many-body spectra near the critical point and in the deep topological regime, compared with results from two low-energy continuum models, respectively. It is clear that both continuum theories show great agreement with solutions from the numerical lattice model in the single-particle spectrum [Figs.~\ref{Fig:Spectrum}(a) and \ref{Fig:Spectrum}(b)], which can be interpreted as follows: When the spin chain is near the critical point $\Omega\rightarrow0$ with $\Gamma\gtrsim\Omega$, the energy gap $2J\left|2t-|g|\right|$ will always happen around $k=0$, where the long wavelength continuum theory dominates. While, if the chain is in a deep topological regime with $\Gamma\ll\Omega$, the spectrum is gapped with $\epsilon_\ssub{gap}\approx4J\gamma$ near the two Fermi points $\pm k_\textsc{f}$ which is in agreement with the deep topological continuum theory. From the perspective of fermionic language, the superconducting coherence length is defined as $\xi=\Upsilon/\Delta=ta/\gamma$ \cite{Kopnin2001}, while the continuum theory requires the coherence length to be much larger than the wave length, i.e., $\xi\gg2\pi/k_\textsc{f}$, which also leads to the validity condition $\gamma\ll t$.

In spite of the excellent agreement between the numerical and analytical results in the single-particle spectra [Figs.~\ref{Fig:Spectrum}(a) and \ref{Fig:Spectrum}(b)], there is only fair agreement between the numerical and analytical results in the many-body spectra [Figs.~\ref{Fig:Spectrum}(c) and \ref{Fig:Spectrum}(d)], where we have globally shifted the energies to make the ground-state energy zero at $\phi=0$. Since the many-body spectra of the low-energy continuum theory can only be constructed by a few single-particle energies of the ABSs in the gap, the contributions from the propagating states outside the gap will not be captured in the analytical continuum theory, which could also lead to small discontinuities in slope at $\phi=l\pi$, $l\in \mathbb{Z}$. Yet, we note that the spectra near the critical point match better than that in the deep topological regime due to the weaker $\phi$-dependence of the propagating state energies.

\section{Fractional Spin Josephson Effect}
\label{Sec:FSJE}

Historically, the original JE was used to describe the supercurrent through a weak link between the conventional $s$-wave superconductors, following $2\pi$ periodicity of the system Hamiltonian \cite{Martinis2004}. Nevertheless, JJs between topological $p$-wave superconductors are predicted to exhibit a $4\pi$-periodic supercurrent, a hallmark manifestation for the existence of MBSs \cite{DengS2016,LiuPRB2017,ZhangN2018}. Notably, a variety of JEs can be identified by coupling the edges of QSH insulators to $s$-wave superconductors. Under the TRS and parity conservation, a dc voltage bias gradually connects the in-gap states to the bulk of scattering states, generating a $2\pi$-periodic dissipative current. Once the TRS is broken, the current becomes dissipationless and evolves as $4\pi$ periodicity, as protected by the PHS stemming from the MBSs \cite{FuPRB2009}. Furthermore, given the TRS with the Coulomb interactions \cite{ZhangPRL2014} or the impurities \cite{PengPRL2016,Vinkler-AvivPRB2017,HuiPRB2017}, the current can even be dissipationless with $8\pi$ periodicity, while the $s_z$-conserving interactions will lead to dissipation with the original $2\pi$ periodicity (note that $s_z$ refers to the electron spin at the QSH edge, instead of the spin in the IXI chain, see Ref.~\cite{HuiPRB2017}). Such $4\pi$ ($8\pi$) periodicity is called $\mathbb{Z}_2$ ($\mathbb{Z}_4$) fractional JE for the sake of $e$ ($e/2$) electron charge being transferred in $2\pi$ period of the system Hamiltonian, instead of Cooper pairs $2e$ in the conventional superconductors. However, in Ref.~\cite{LaflammeNC2016} it was shown that such $8\pi$ periodicity can be achieved \textit{without} Coulomb interactions, based on a $p$-wave superconductor lattice ring interrupted by one weakly coupled normal site. 

Before analyzing the spin JEs in our setup, we want to make a key observation: The spin twisting angle $\phi$ has been mapped into the superconducting phase $2\phi$, i.e., it was doubled, which makes all periodicities of the fermionic JEs twice as large as the spin JEs. Explicitly, the periodicities of trivial, $\mathbb{Z}_2$, $\mathbb{Z}_4$ JEs become $\pi$, $2\pi$, $4\pi$ in the spin chain, respectively, compared with $2\pi$, $4\pi$, $8\pi$ in the fermionic systems. To avoid confusion, in the following discussions, we will use trivial, $\mathbb{Z}_2$, $\mathbb{Z}_4$ terms to illustrate various JEs in the two representations.

Although the properties of fractional JEs in the fermionic systems are well-studied, a question naturally arises: Except for the alteration at the phase $\phi$ by a factor of 2, what are the similarities and differences between fermionic JEs and spin JEs? In the following subsections, we will investigate various spin JEs from two perspectives: the continuum theory and the lattice model. Moreover, to reveal the influence of the many-body interactions on the spin fractional JEs, we will add ZZ-type interactions [Eq.~\eqref{Eqn:HSpinZZ}] and NN-type interactions [Eq.~\eqref{Eqn:HFermiNN}] into Eqs.~\eqref{Eqn:HSpinGeneral} and \eqref{Eqn:HFermiGeneral}, respectively, both of which act only within the middle sector. We note that these interactions, which are quartic in fermionic operators, force us to apply a brute-force diagonalization on a $2^N\times 2^N$ matrix in spin space, effectively limiting the number of sites, $N$, of the chain.

\subsection{Continuum Scenarios}

\begin{figure}
  \centering
  \includegraphics[width=0.48\textwidth]{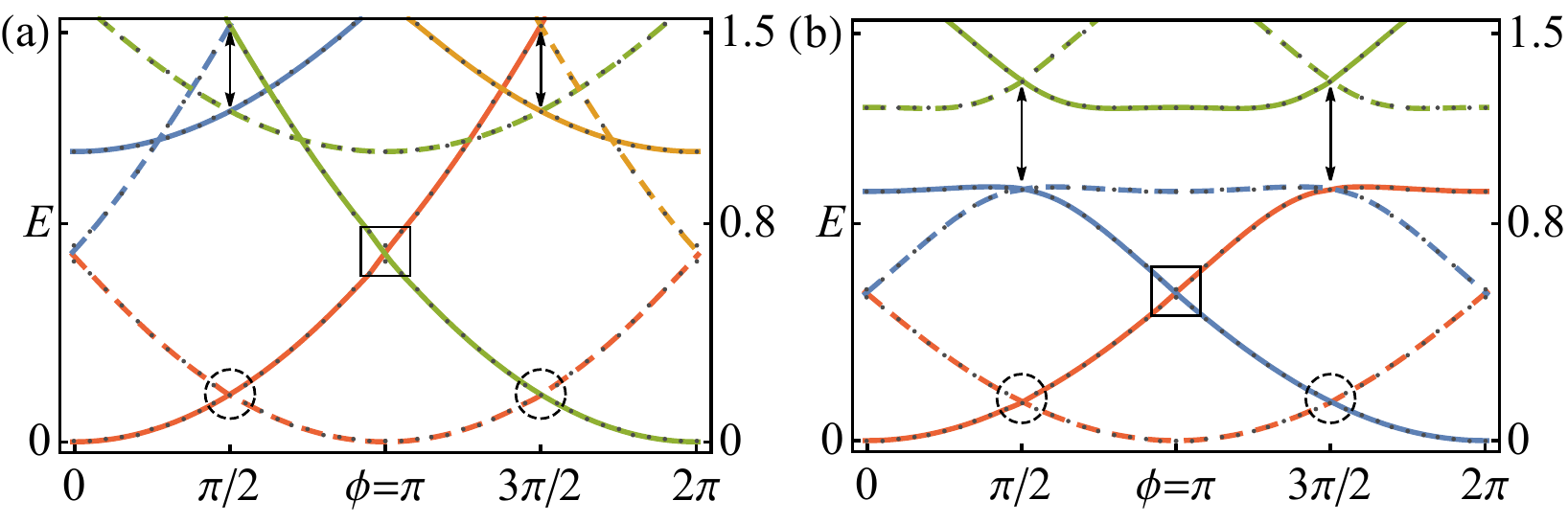}
  \caption{Many-body spectra of the IXI varied as $\phi$, calculated by the exact diagonalization under $J=t=\mathbbm{t}=1$, $\gamma=0.6$, $g=0$, $N_\textsc{l}=N_\textsc{r}=9$, $N_\textsc{m}=6$ after adding two types of interactions. (a) is under ZZ-type (spin) interactions [Eq.~\eqref{Eqn:HSpinZZ}] with $\delta=0.4$, while (b) includes NN-type (fermionic) interactions [Eq.~\eqref{Eqn:HFermiNN}] with $\chi=0.4$. Solid (dashed) lines indicate even (odd) parity supplemented with original data (black dots), dashed circles are crossings protected by the PHS, rectangles refer to crossings protected by the eTRS in the continuum limit while broken by finite-size effects, gaps at the arrows are lifted by interactions. }
  \label{Fig:ContiZZNN}
\end{figure}

In the low-energy continuum limit, both Eqs.~\eqref{Eqn:HCriticalGeneral} and \eqref{Eqn:HDeepGeneral} obey PHS: $\{\mathcal{H}_\textsc{g}^\textsc{c}, \mathcal{C}_\textsc{c}\}=0$, $\mathcal{C}_\textsc{c}=\rho_x \mathcal{K}$ near the critical point and $\{\mathcal{H}_\textsc{g}^\textsc{d}, \mathcal{C}_\textsc{d}\}=0$, $\mathcal{C}_\textsc{d}=\rho_y \tau_y \mathcal{K}$ in the deep topological regime, which guarantees the crossings of MBSs and switches the parity of the ground state at $\phi=\pi/2 + l\pi$. Additionally, as we have shown in Sec.~\ref{Sec:LowEnergy}, crossings at $\phi=l \pi/2$ are protected by the eTRS of Eq.~\eqref{Eqn:HDeepGeneral} in the deep topological regime, which is indeed equivalent to JJs attached to the edge of QSH insulators. Therefore, adiabatically advancing the spin twisting angle $\phi$ will pump each ABS into the bulk and lead to dissipative current with trivial periodicity, as displayed in Figs.~\ref{Fig:Spectrum}(b) and \ref{Fig:Spectrum}(d). Nonetheless, when the system is tuned close to the critical point where the eTRS is broken, there are anti-crossings at $\phi=l \pi/2$ in Figs.~\ref{Fig:Spectrum}(a) and \ref{Fig:Spectrum}(c), with the exception of the low-energy crossings (dashed circles) at $\phi=\pi/2 + l\pi$ that are still protected by the Majorana PHS. Under this circumstance, every ABS is detached from the bulk and give rise to dissipationless spin current with $\mathbb{Z}_2$ periodicity. 

In Fig.~\ref{Fig:ContiZZNN}, we show the many-body spectra in the deep topological regime, taking into account interactions of ZZ type [Eq.~\eqref{Eqn:HSpinZZ}] and NN type [Eq.~\eqref{Eqn:HFermiNN}], respectively, both still with the eTRS maintained. Compared with Fig.~\ref{Fig:Spectrum}(d), prior fourfold degeneracy at $\phi=\pi/2$ is lifted via the Coulomb interactions [indicated by the vertical arrows in Fig.~\ref{Fig:ContiZZNN}(b)], a dissipationless $\mathbb{Z}_4$ spin current occurs as expected \cite{ZhangPRL2014}. Conversely, ZZ-type interactions only shift crossings [indicated by the vertical arrows in Fig.~\ref{Fig:ContiZZNN}(a)]. Because the energy levels move into the bulk as $\phi$ is increased, the spin current remains dissipative with trivial periodicity as in the aforementioned non-interacting case. This phenomenon basically resembles QSH JJs accompanied with $s_z$-conserving interactions in Ref.~\cite{HuiPRB2017}. Although there are small gaps at $\phi=\pi$ caused by finite-size effects (e.g., slowly oscillatory umklapp or Friedel terms), they can be fairly suppressed under the continuum limit \cite{LopesPRB2019}.

\subsection{Lattice Odd-Even Effect}
\label{Sec:OddEven}

\begin{figure}
  \centering
  \includegraphics[width=0.48\textwidth]{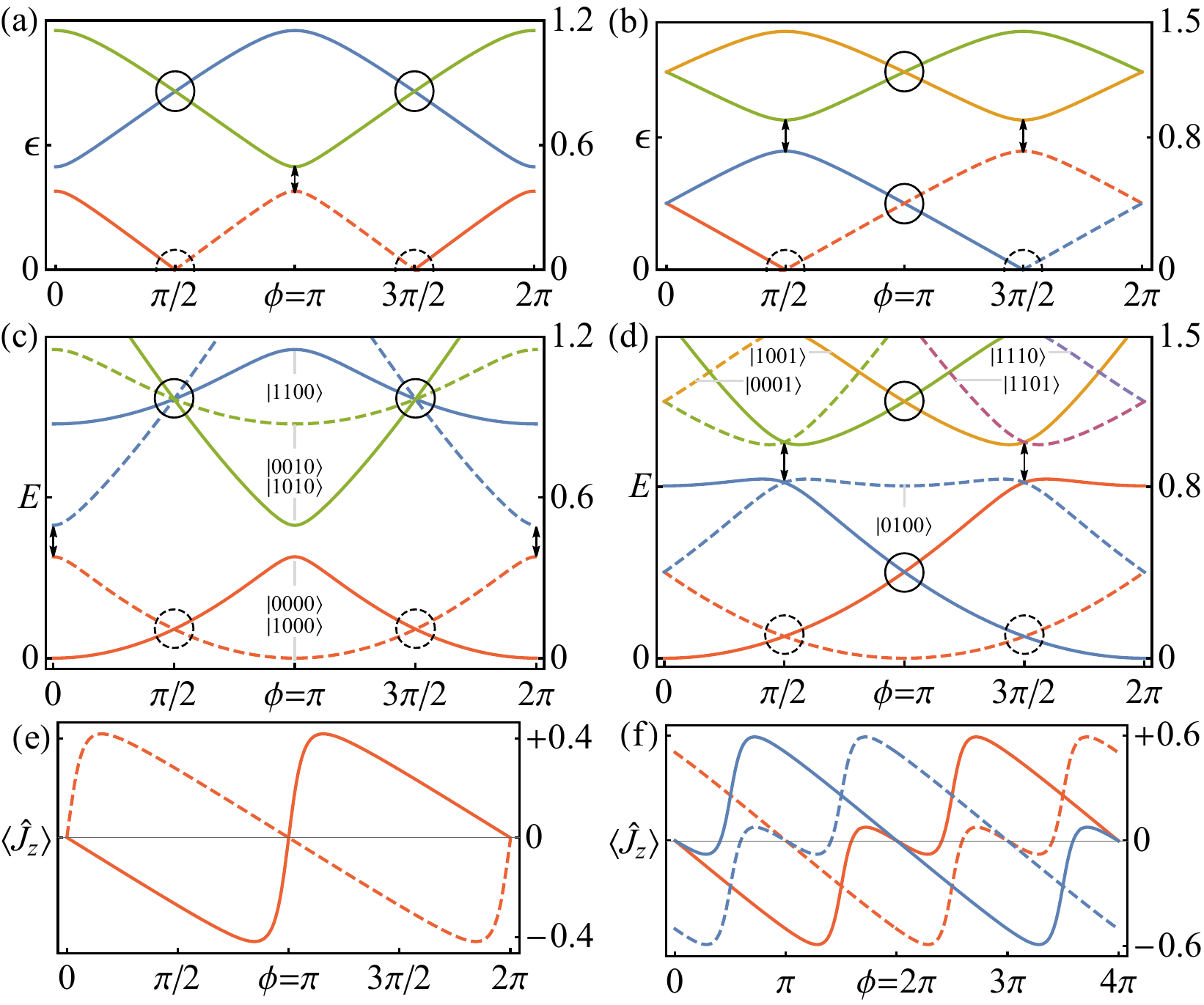}
  \caption{The odd-even effect: spectra and spin supercurrent of the IXI as a function of $\phi$, solved numerically by the BdG matrix diagonalization under $J=t=1$, $\mathbbm{t}=0.8$, $\gamma=0.4$, $g=0$, $N_\textsc{l}=N_\textsc{r}=100$ in both odd-even cases. (a) and (b) are the single-particle spectra for $N_\textsc{m}=10$ and $N_\textsc{m}=11$, respectively. (c) and (d) are their corresponding many-body spectra, whose single-particle occupations are shown in plot labels. (e) and (f) are spin current of their lowest two and four states evaluated from Eq.~\eqref{Eqn:SpinCurrent}, where the variation of $\phi$ is extended to full $4\pi$ period in (f), showing $\mathbb{Z}_2$ and $\mathbb{Z}_4$ periodicities respectively. Solid (dashed) lines in the single-particle spectra are the energies of the particles (holes), solid (dashed) lines in the many-body spectra refer to the even (odd) parity, solid (dashed) circles are crossings protected by the iTRS (PHS), the gaps specified by the arrows are lifted by the imperfect couplings $\mathbbm{t}<t$.}
  \label{Fig:LatticeOddEven}
\end{figure}

The eTRS in the continuum limit requires the transport through JJs to be highly transparent, any imperfect connections $\mathbbm{t}\neq t$ are able to break such symmetry and open gaps at the lattice level, which leads to the following odd-even effects. As we have proven in Sec.~\ref{Sec:Symmetry}, there is an iTRS appearing at the lattice level when all parameters are set inverted symmetrically, bringing about different crossing properties for odd-even sites. In particular, for all single-particle states illustrated in Figs.~\ref{Fig:LatticeOddEven}(a) and \ref{Fig:LatticeOddEven}(b), there must be Kramers pairs at $\phi=l\pi$ for odd $N$ and $\phi=\pi/2+l\pi$ for even $N$, according to the conclusions of Eq.~\eqref{Eqn:InversionTRSphi}. By changing the parity of the sites, crossings and anti-crossings can be created or destroyed at specific $\phi$ in the spectra, shown in Fig.~\ref{Fig:LatticeOddEven}. As a consequence, adiabatically following the ground states will eventually lead to $\mathbb{Z}_2$ ($\mathbb{Z}_4$) spin current for the even (odd) sites, pumping different amounts of net spin between the left and the right Ising parts, as displayed in Figs.~\ref{Fig:LatticeOddEven}(e) and \ref{Fig:LatticeOddEven}(f) calculated by Eq.~\eqref{Eqn:SpinCurrent} [or Eq.~\eqref{Appx:SpinCurrent}, see Appendix~\ref{Appx:CoFunc} for details]. Alternatively, because there are no many-body interactions, the spin current can be analytically computed using $\braket{\hat{J}_z}_n=-2\partial E_n/\partial \phi$, upon applying a phase-shifted JWT $\ch^\dagger_i = \expm^{-\im \phi} \prod_{j=1}^{i-1} (-\sigmaz_j) \sigmap_i$ on the right part and transforming $\phi$ into the right interface \cite{TserkovnyakPRA2011}, which gets along with conventional results for the fermionic Josephson current \cite{Martinis2004}. To evaluate the full adiabatic spin current in the presence of a time-dependent angle twist $\phi(t)$, one needs to account for the possible Berry phase contributions to the current stemming from the velocity of the twist, $\dot{\phi}(t)$ which, however, is beyond the scope of this work \cite{MottonenPRL2008}.

In addition, our conclusion reveals the unusual $\mathbb{Z}_4$ fractional JE in Ref.~\cite{LaflammeNC2016} is actually protected by the iTRS. In fact, their model Hamiltonian is equivalent to ours for $N_\textsc{m}=1$ after applying the phase-shifted JWT \footnote{Although periodic boundary conditions are imposed in their $p$-wave superconducting parts to form a ring geometry, degenerate properties within the gap are still well established.}. The reason why in their case the $\mathbb{Z}_4$ periodicity cannot survive under the Coulomb interactions is that NN-type interactions do not commute with iTRS, whereas ZZ-type interactions do, as it happens in spin chains \cite{Giamarchi2003}. Namely, the spectra may be shifted under ZZ-type interactions while crossings are still protected. Therefore, $\mathbb{Z}_4$ spin current originating from iTRS does not depend on whether there are ZZ-type interactions or not.

\section{Texture of Spin Entanglement}
\label{Sec:Entanglement}

\begin{figure}
  \centering
  \includegraphics[width=0.48\textwidth]{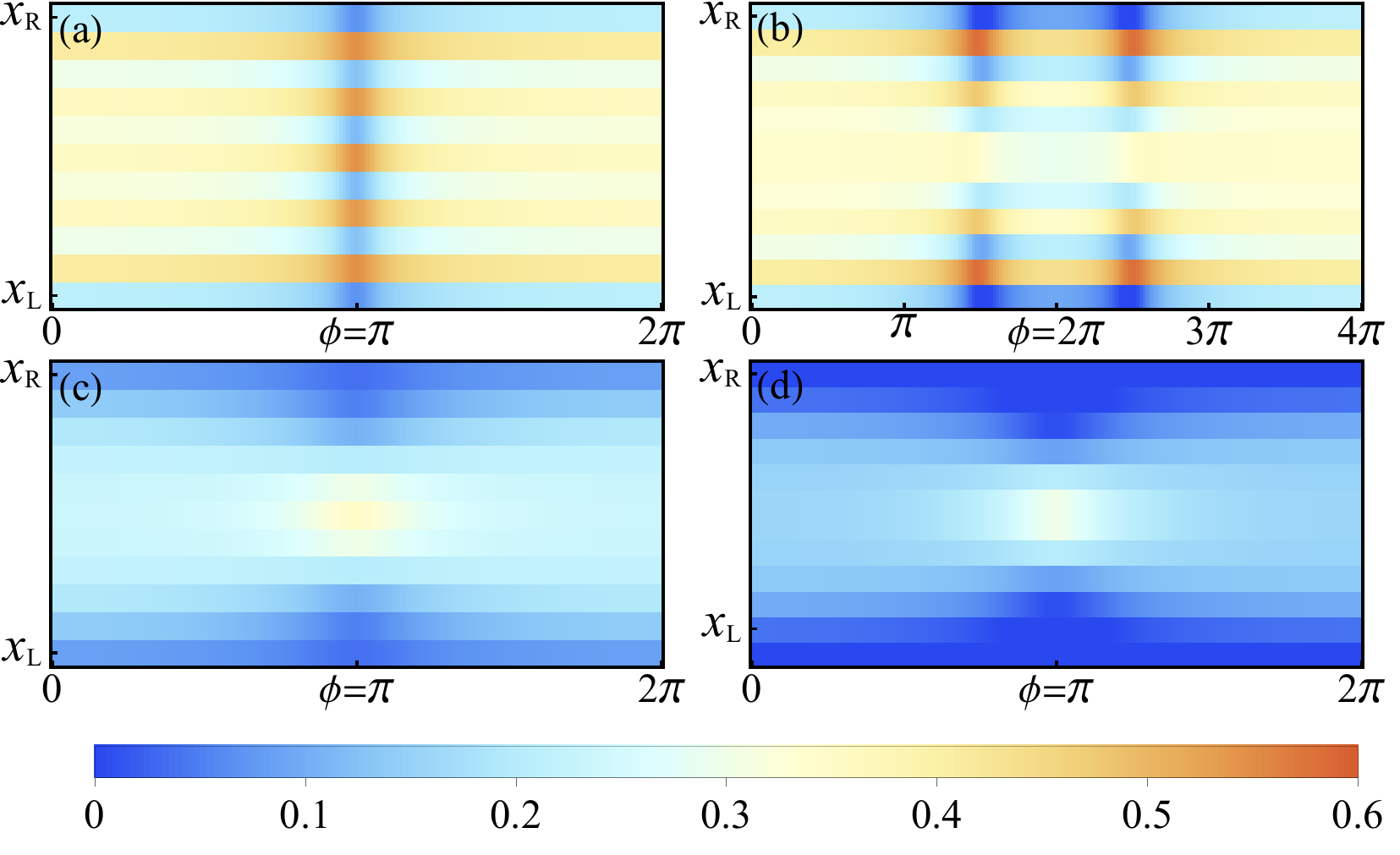}
  \caption{Concurrences in the XY sector for ground states as a function of $\phi$, computed from the Pfaffian of the correlation matrices under $J=t=1$, $\mathbbm{t}=0.8$, $\gamma=0.4$, $N_\textsc{l}=N_\textsc{r}=100$ in all sub-figures. (a) and (b) are nearest-neighbor concurrences in the deep topological regime $g=0$ for $N_\textsc{m}=10$ and $N_\textsc{m}=11$ separately, where the variation of $\phi$ is extended to full $4\pi$ period in (b). While (c) and (d) are nearest-neighbor and next-nearest-neighbor concurrences for $N_\textsc{m}=10$ near the critical point $g=-1.8$, respectively.}
  \label{Fig:Concurrence}
\end{figure}

In this section, we evaluate various spin correlation functions in the presence of the spin supercurrent carried by JW Majoranas in the XY sector. Specifically, we are interested in the single spin expectation value $p_i^\alpha \equiv p_i^\alpha(\phi) = \braket{\sigmah_i^\alpha}$, as well as the spin-spin correlation function $p_{ij}^{\alpha\beta} \equiv p_{ij}^{\alpha\beta}(\phi) = \braket{\sigmah_i^\alpha\sigmah_j^\beta}$ with $\alpha,\beta=x,y,z$. This allows us to derive the reduced density matrices for an arbitrary single and pair of spins,
\begin{equation}
\rho_i(\phi)=\frac{1}{2}\sum_{\alpha=0}^3p_i^\alpha\sigmah_i^\alpha,\quad \rho_{ij}(\phi)=\frac{1}{4}\sum_{\alpha,\beta=0}^3p_{\alpha\beta}\sigmah_i^\alpha\sigmah_j^\beta\,,
\end{equation}
respectively. Since the Hamiltonian conserves the parity of the system, we can readily infer that $p_i^x=p_i^y=0$, thus the spin texture has only one non-zero component $p_i^z$, along the $z$-direction. Similarly for the two-spin correlators, several components vanish: $p_{ij}^{xz}=p_{ij}^{zx}=p_{ij}^{yz}=p_{ij}^{zy}=0$. It is clear from the definition of spin current [Eq.~\eqref{Eqn:SpinCurrent}] that when there is a finite spin supercurrent flowing through the middle part, $p_{ij}^{xy}$ and $p_{ij}^{yx}$ must be nonzero. In this case, regular determinant stratagems \cite{LiebAP1961,OsbornePRA2002,Sachdev2011} cannot be used to find the correlator between two arbitrary spins. However, such correlators, together with nonzero $p_{ij}^{xx}$ and $p_{ij}^{yy}$,  can be obtained by computing the Pfaffian of their corresponding $2k\times2k$ skew-symmetric matrices \cite{CaianielloNC1952,BarouchPRA1971}, where $k=|i-j|$ (see Appendix~\ref{Appx:CoFunc}).

With all spin correlators at hand, we are able to establish the reduced density matrices, and then evaluate the degree of entanglement in the system. There are two simple subsystems in which one can easily calculate the entanglement \cite{OsbornePRA2002}: (1) a single site and the rest of the lattice and  (2) two arbitrary spins in the chain. For the former, the entanglement can be calculated via the \textit{von Neumann entropy} $S_i(\phi)=-\mathrm{tr}[\rho_i(\phi)\log{\rho_i(\phi)}]$, assuming the whole chain in a pure state. For the two sites case in a mixed state, the amount of entanglement shared between the spins is quantified by the concurrence $C$. In particular, for two arbitrary spin-$\nicefrac{1}{2}$ sites at the positions $i$ and $j$ in the chain, the concurrence is given by \cite{WoottersQIC2001}
\begin{equation}
C(\rho_{ij})={\rm max}[0,\lambda_{ij}^1-\lambda_{ij}^2-\lambda_{ij}^3-\lambda_{ij}^4]\,,
\end{equation}
where the $\lambda_{ij}^k$ are the eigenvalues of the Hermitian matrix $R_{ij}=\sqrt{\sqrt{\rho_{ij}}\tilde{\rho}_{ij}\sqrt{\rho_{ij}}}$ sorted in descending order with $\tilde{\rho}_{ij}=(\hat{\sigma}_i^y\otimes\hat{\sigma}_j^y)\rho_{ij}^*(\hat{\sigma}_i^y\otimes\hat{\sigma}_j^y)$. The concurrence increases from $C=0$ for a separable state to $C=1$ for a maximally entangled state. Although the single-site entropy and the concurrence between two arbitrary spins is known to peak at the quantum phase transition \cite{OsbornePRA2002}, here we determine how the entanglement in the XY sector is affected by the presence of spin supercurrent due to a finite twist between the Ising directions.

In Fig.~\ref{Fig:Concurrence} we plot the texture of the spin concurrences as a function of $\phi$ for odd-even cases in different regimes, following the ground states in Fig.~\ref{Fig:LatticeOddEven}. It is apparent to see that there are two different textures of spin entanglement for odd-even cases depicted in Figs.~\ref{Fig:LatticeOddEven}(a) and \ref{Fig:LatticeOddEven}(b), not only evolving with two kinds of periodicities, but also taking peaks (nadirs) at different $\phi$. Such phenomena are due to the fact that through increasing $\phi$, the many-body levels have been shifted to higher values, which makes them more susceptible to higher excited states. Owing to finite size effects with open boundary conditions, the entanglement also oscillates with frequency $\sim2 k_\textsc{f}$ as a function of site index \cite{CalabresePRL2010}, which can be enhanced by larger susceptibilities close to anti-crossing points. Hence, one can strongly control the entanglement between the spins in the XY sector via the twisting angle, which could be utilized to process quantum information. 

Furthermore, by comparing Figs.~\ref{Fig:LatticeOddEven}(a) and \ref{Fig:LatticeOddEven}(b) to \ref{Fig:LatticeOddEven}(c) and \ref{Fig:LatticeOddEven}(d), one might wonder why concurrences near the critical point are less than that in the deep topological regime, since the chain should be more entangled around quantum phase transition. The reason is as follows: In the deep topological regime, only nearest-neighbor concurrences are nonzero, which means the entanglement is well confined in nearest-neighbor spins; while as the system approaches the critical point, the entanglement will be spread out into next-nearest-neighbor (and so on) spins \cite{OsbornePRA2002}, which makes the initial nearest-neighbor concurrence decrease.

\section{Detection and Robustness}
\label{Sec:cQED}

In this section, we address the detection of the spin supercurrent pertaining to the JW Majoranas in the IXI spin junction. While the method of choice for measuring spin current is through the use of the spin Hall effect \cite{SinovaRMP2015}, in which case a spin current is converted to a charge current that can be measured by usual techniques, via the SOI in the adjacent material, here we propose a less invasive method based on microwave detection. Such an approach has been found suitable for measuring both the statics and dynamics of ABSs in electronic systems \cite{DassonnevillePRL2013,MuraniPRL2019,AftergoodPRB2019}. The idea is to couple the field of a nearby resonator to various observables of the system. The interaction between our chain and the resonator can be written as
\begin{equation}
\hat{V}(t)=\beta\hat{O}(a^\dagger+a)\,,
\end{equation}
where $a$ ($a^\dagger$) is the annihilation (creation) operator for the photon in the resonator (assuming one mode only), while $\hat{O}$ are the observables of the system, e.g., $\hat{O}=\sigmah^\alpha_i$ (or the sum of a string of spins), with coupling strength $\beta$. This coupling will alter the properties of the resonator, which in turn can be measured in a dispersive readout. Following Ref.~\cite{DmytrukPRB2015}, we can write the equation of motion for the cavity field in the Heisenberg picture as
\begin{equation}
    \dot{a}=\im[\hat{H}_\mathrm{ph}+\hat{V}(t),a]-\frac{\kappa}{2}a-\sqrt{\kappa}b_\mathrm{in}(t)\,,
\end{equation}
where $\hat{H}_\mathrm{ph}=\omega_0 a^\dagger a$ is the cavity Hamiltonian, $\kappa$ quantifies the decay rate of the cavity, and $b_\mathrm{in}(t)$ is the input field sent to probe it. Note that the output field, exiting from the cavity $b_\mathrm{out}(t)$, and the input one satisfy $b_\mathrm{out}(t)=b_\mathrm{in}(t)+\sqrt{\kappa}a(t)$, which is used to infer the cavity response. In leading order in the cavity-system coupling and in the frequency space, we find \cite{DmytrukPRB2015}
\begin{equation}
    a(\omega)=-\frac{\sqrt{\kappa}b_\mathrm{in}(\omega)+\im\beta\langle \hat{O}_\textsc{i}(\omega)\rangle_0}{-\im(\omega-\omega_0)+\kappa/2-\im\,\beta^2\Pi_\ssub{\hat{O}}(\omega_0)}\,,
\end{equation}
where $a(\omega)=\int \dm t\, \expm^{-\im \omega t}a(t)$ and
\begin{align}\label{Eqn:SSMany}
    \Pi_\ssub{\hat{O}}(\omega)&=-\im\int_0^\infty \dm t\, \expm^{-\im \omega t} \langle[\hat{O}_\textsc{i}(t),\hat{O}_\textsc{i}(0)]\rangle_0\nonumber\\
    &=\sum_{m,n}'\frac{|\langle m|\hat{O}|n \rangle|^2(F_m-F_n)}{E_m-E_n-\omega-\im\eta}\,,
\end{align}
being the retarded correlation function associated with the observable $\hat{O}$ over the stationary state of the system $\langle\dots\rangle_0$. Above, $|n\rangle$ and $E_n$ are the many-body eigen-states and eigen-energies of the system, respectively, $F_n$ is the many-body occupation, while the $'$ index selects only the states $n\neq m$ in the summation. Note that all quantities are expressed in the interaction picture, and $\langle \hat{O}_\textsc{i}(\omega)\rangle_0$ is the expectation value of the observable $\hat{O}$ in the frequency space in the absence of the cavity. Since the energies $E_n$, as well as the matrix elements $\langle m|\hat{O}|n\rangle$ are functions of $\phi$, the entire correlation function will carry such a dependence too. In typical spectroscopic experiments, the input field $b_\mathrm{in}(\omega)\gg\langle\hat{O}_\textsc{i}(\omega)\rangle_0$ (large number of photons are sent into the cavity), and we can neglect this term in the following. Nevertheless, such contribution can become relevant in out-of-equilibrium situations, when it affects the photon number and photon statistics in the cavity. We will not discuss such regimes here, but refer to Ref.~\cite{AftergoodPRB2019} for some details (along with the schematic of cQED setups). The effect of the spins on the cavity photons results in changes in both the resonance frequency  $\omega_0$ and the quality factor (or $Q$ factor) of the cavity, which can be straightforwardly related to the correlation function as follows: 
\begin{align}
    \delta\omega_0(\phi)&=\beta^2{\rm Re}\;\Pi_\ssub{\hat{O}}(\omega_0,\phi)\,,\\
    \frac{\delta Q(\phi)}{Q}&=\frac{\beta^2{\rm Im}\;\Pi_\ssub{\hat{O}}(\omega_0,\phi)}{\omega_0}\,,
\end{align}
implying quadratic dependence on the coupling strength $\beta$ of these quantities. This coupling depends on the specific implementation of our model, ranging from a tens of Hz for electron spins coupled directly to the magnetic component of an electromagnetic cavity, to tens of MHz  in the case of superconducting qubits (in which case the coupling occurs via the electrical field of the cavity instead).  

In this paper, we consider a capacitive-like coupling between the spin chain and the cavity \textit{magnetic} field (through the Zeeman coupling), following Ref.~\cite{AftergoodPRB2019}. Moreover, we assume the \textit{magnetic} field of a microwave cavity couples to the spins in the XY part over a length $l<L$, or $\hat{O}=\bm{\hat{S}}_l\cdot\mathbf{n}$, with $\bm{\hat{S}}_l=\sum_{i\in l}\bm{\sigmah}_i$. Here, $\mathbf{n}$ is the direction of the cavity magnetic field at the position of the wire, which can be different from the $z$ direction, and the coupling is assumed to take place from site $l_0$ to site $l_0+l-1$. The susceptibility can be written as $\Pi_\textsc{s}(\omega)=\Pi_\textsc{s}^z(\omega)+\Pi^\ssub{\perp}_\textsc{s}(\omega)$, where the first and second terms corresponding to the matrix element $\langle m|\hat{S}^z_l|n\rangle$ (longitudinal) and $\langle m|\bm{\hat{S}}_l\cdot\mathbf{n}_\ssub{\perp}|n\rangle$ (transverse), respectively, with $\mathbf{n}_\ssub{\perp}=\mathbf{n}-\mathbf{e}_z$. There are no cross terms between the $z$ (parity preserving) and $x,y$ (parity flipping) spin components as all the states in the system have a definite parity. The above susceptibilities have a simple interpretation in the fermionic language: The first contribution stems from the cavity probing particle number operator over the length $l$, while the second one effectively represents electronic tunneling into the spin chain over the same distance, thus accessing the transport properties of the spin chain. However, as we see in the following discussions, the analogy is only partial for the second coupling because of the non-locality of the JW string.

\begin{figure}
  \centering
  \includegraphics[width=0.48\textwidth]{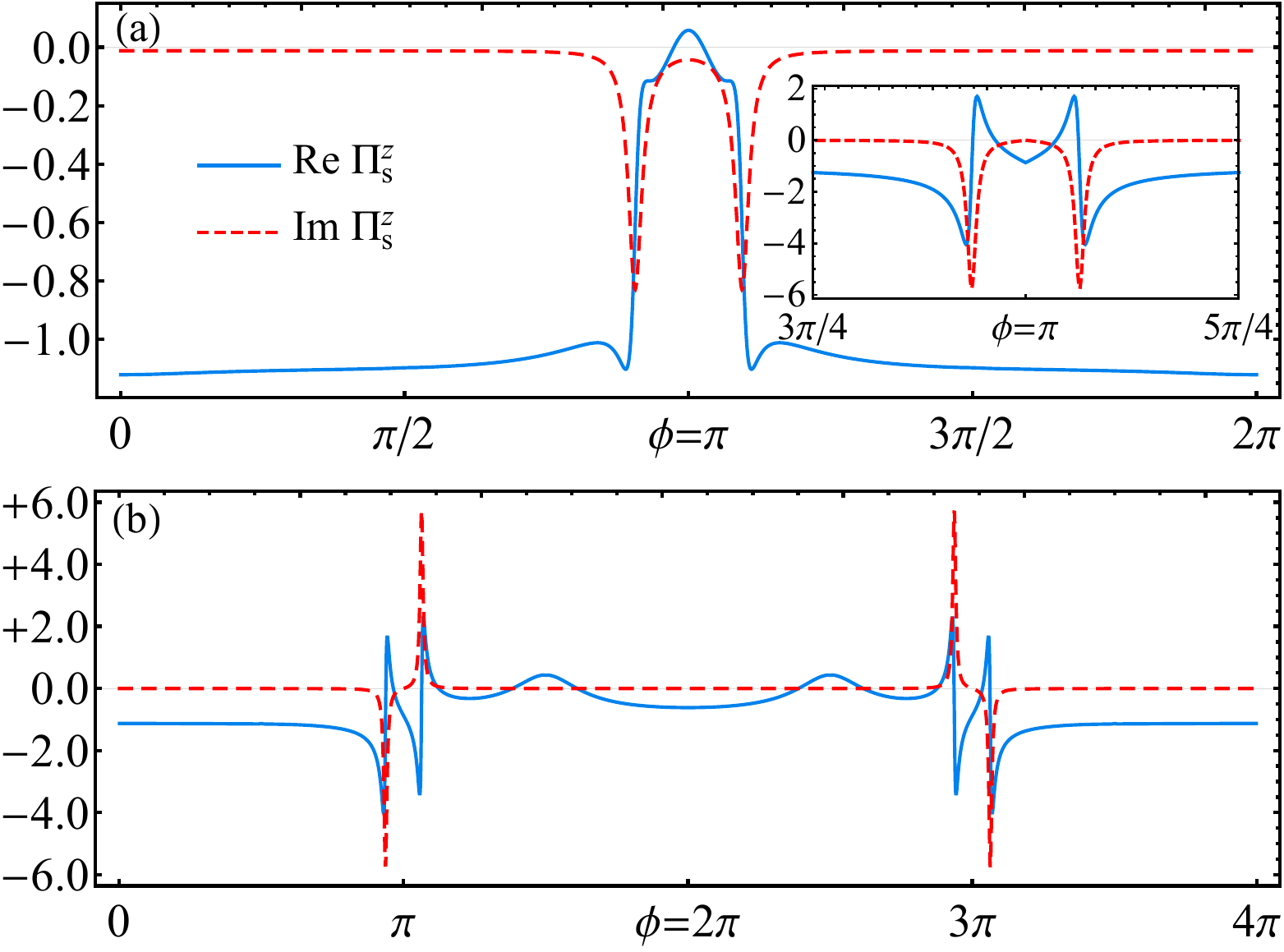}
  \caption{Dependence of the longitudinal susceptibility on $\phi$ calculated by Eq.~\eqref{Eqn:SSsingle} with $J=t=1$, $\mathbbm{t}=0.8$, $\gamma=0.4$, $g=0$, $\eta=0.1\omega$, $N_\textsc{l}=N_\textsc{r}=100$, whose real (imaginary) parts are represented by blue (red dashed) lines. In both (a) and (b), the cavity couples to five spins starting from $l_0=N_\textsc{l}+3$. More specifically, (a) is $\mathbb{Z}_2$ case with $N_\textsc{m}=10$, $\omega=0.2$, which shows negative peaks of the imaginary parts around $\phi=\pi$; (b) is $\mathbb{Z}_4$ case with $N_\textsc{m}=11$, $\omega=0.1$, which shows peaks of the imaginary parts around $\phi=\pi$ that have opposite signs. In the presence of relaxation, for odd number of sites, the susceptibility will return to $\mathbb{Z}_2$ periodicity as shown in the inset of (a), which exhibits a singularity in the real part at $\phi=\pi$.}
  \label{Fig:Susceptibility}
\end{figure}

\subsection{Longitudinal Susceptibility}

The longitudinal susceptibility can now be numerically evaluated from the lattice model by including all possible states. However, in order to understand the behavior, it is worth analyzing the limit of small $\omega\ll\Delta$ in which case the cavity probes mostly the low-energy ABSs (truncated up to the 12th state in calculation), including the MBSs. We transform the spins into fermions in the lattice $\ch_i$, and eventually in terms of quasi-particles describing the Andreev states $\dch_n$, with $i$ and $n$ specifying the lattice and eigen-energy index, respectively. By using $\ch_i = \sum\nolimits_n [ u_n(i) \dch_n + v^*_n(i) \dch^\dagger_n ]$ with coefficients $u_n(i)$ and $v_n(i)$ found from wave functions of numerical diagonalization (see Appendix~\ref{Appx:SpinChain} for details), we write down $\hat{S}^z_l$ in the form of quasi-particles, 
\begin{equation}
\hat{S}^z_l
=\sum_{i\in l}\sum_{r,s} [ b^*_r(i) \dch^\dagger_r - b_r(i) \dch_r ] [ a^*_s(i) \dch^\dagger_s + a_s(i) \dch_s ]\,,
\end{equation}
with $a_s(i)=u_s(i) + v_s(i)$, $b_s(i)=u_s(i) - v_s(i)$, where $r,s$ are single-particle indices of their corresponding many-body states in Eq.~\eqref{Eqn:SSMany}, given in the labels of Figs.~\ref{Fig:LatticeOddEven}(c) and \ref{Fig:LatticeOddEven}(d). There are two types of $\braket{m|\hat{S}^z_l|n}$: quasi-particle conserving type $S^\mathrm{c}_{r,s}$ and non-conserving type $S^\mathrm{n}_{r,s}$, which are shown explicitly as
\begin{align}
S^\mathrm{c}_{r,s}&=\sum_{i\in l} [b^*_r(i) a_s(i) + b_s(i) a^*_r(i)] \,,\\
S^\mathrm{n}_{r,s}&=\sum_{i\in l} [b_r(i) a_s(i) - b_s(i) a_r(i)]\,.
\end{align}
With single-particle occupation $f_s\equiv\braket{\dch^\dagger_s \dch_s}$, the longitudinal susceptibility is written in the single-particle form:
\begin{align}\label{Eqn:SSsingle}
&\Pi^z_\textsc{s}(\omega)=\sum_{r,s}'\left[\frac{(f_r-f_s)|S^\mathrm{c}_{r,s}|^2}{\epsilon_r-\epsilon_s-\omega-\im\eta}+\frac{(f_r-f_s)|S^\mathrm{c}_{r,s}|^2}{\epsilon_r-\epsilon_s+\omega+\im\eta}\right.\nonumber\\
&+\left.\frac{(f_r+f_s-1)|S^\mathrm{n}_{r,s}|^2}{\epsilon_r+\epsilon_s-\omega-\im\eta}+\frac{(f_r+f_s-1)|S^\mathrm{n}_{r,s}|^2}{\epsilon_r+\epsilon_s+\omega+\im\eta}\right],
\end{align}
where the first (second) line accounts for the quasi-particle conserving (non-conserving) contributions. 

In Figs.~\ref{Fig:Susceptibility}(a) and \ref{Fig:Susceptibility}(b), we show the real and the imaginary parts of $\Pi^z_\textsc{s}(\omega)$ as a function of $\phi$ for odd and even cases, respectively, evolving adiabatically in their initial ground states at $\phi=0$, whose peaks indicate the resonances between the cavity and the low-energy levels in Figs.~\ref{Fig:LatticeOddEven}(c) and \ref{Fig:LatticeOddEven}(d). They present different periodicities and reach peaks at different $\phi$, as a result of the odd-even effect. Particularly, one can distinguish $\mathbb{Z}_4$ spin current from the $\mathbb{Z}_2$ case, by way of opposite signs near $\phi=\pi$ in the imaginary parts. Moreover, even taking into account the relaxation effects such that the system always follows the ground state, the real part still exhibits a singularity at $\phi=\pi$ in Fig.~\ref{Fig:Susceptibility}(c), which is again a signature for $\mathbb{Z}_4$ crossing of the levels. We note that while the magnetic coupling to each individual spin is typically small (a few Hz in cQED setups), by coupling the cavity to many spins in the chains $\bm{\hat{S}}_l$, the response function is enhanced by an order $\sim l^2$ as compared to the single spin scenario.

\subsection{Transverse Susceptibility and Spin Noise}

\begin{figure}
  \centering
  \includegraphics[width=0.48\textwidth]{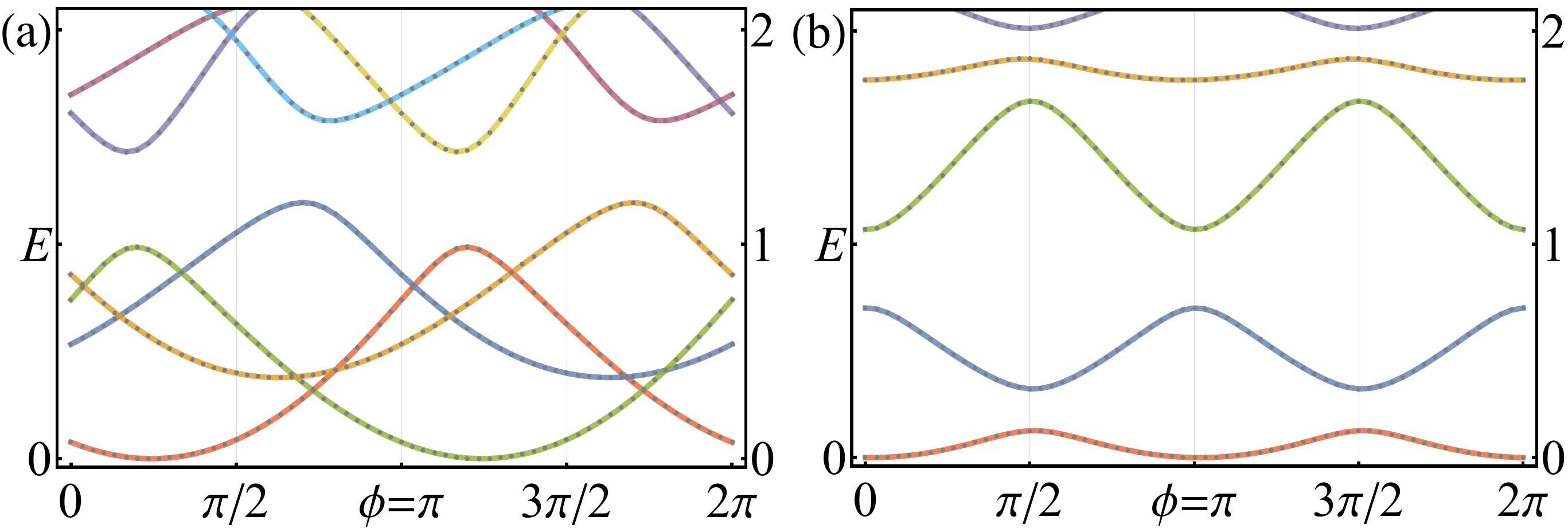}
  \caption{Spectra of the IXI varied as $\phi$ under random perturbations for a given realization, computed by exact diagonalization with $J=t=\gamma=1$, $\mathbbm{t}=0.8$, $g=-0.2$, $N_\textsc{l}=N_\textsc{m}=N_\textsc{r}=4$. (a) is under local spin perturbations [Eq.~\eqref{Eqn:SpinPerturbation}] from the in-plane magnetic fields, crossings are preserved albeit with lifted degeneracies. (b) suffers local fermionic perturbations [Eq.~\eqref{Eqn:FermiPerturbation}] from the quasi-particle poisoning, crossings are destroyed while each state still contains twofold degeneracy. All perturbation strengths $\eta^x_i$ and $\eta^y_i$ are set randomly site by site in the middle XY chain within the range of $(0, 0.2)$. }
  \label{Fig:Robustness}
\end{figure}

\begin{table*}
  \caption{\label{Tab:Summary} A concise summary of conclusions, contrasted with the electronic counterpart.}
  \begin{ruledtabular}
  \begin{tabular}{cccccc}
   & \multicolumn{2}{c}{Fractional Josephson effects} & Entanglement & Detection & Robustness \\
   & Continuum & Lattice &  & & \\ \hline
   SNS& 
   $\mathbb{Z}_2$ or $\mathbb{Z}_4$ & $\mathbb{Z}_2$ or $\mathbb{Z}_4$ & No counterpart   & Transport & No \\
   IXI& 
   $\mathbb{Z}_2$ &  $\mathbb{Z}_2$ or $\mathbb{Z}_4$ & $\phi$-dependence ($\mathbb{Z}_2$ or $\mathbb{Z}_4$) & cQED            & Yes            \\
  \end{tabular}
  \end{ruledtabular}
\end{table*}

Borrowing from the fermionic parity-flipping picture due to the quasi-particle poisoning, one may conjecture that the transverse susceptibility $\Pi^\ssub{\perp}_\textsc{s}(\omega)$ has a nonzero value. Surprisingly, we find out numerically that the matrix elements of $\Pi^\ssub{\perp}_\textsc{s}(\omega)$ are exponentially reduced to \textit{zero} as the length of the Ising part increases, which makes transitions between different parities \textit{impossible} in the topological spin JJs. Such phenomenon is because the local in-plane spin operators $\sigmax_i, \sigmay_i$ become highly non-local objects with the additional JW string in the fermionic space --- it is inevitable to alter the states of \textit{external} JW Majoranas, which in turn flips the parity back to itself and thus forbids the transitions between them. 

To verify this, we study the influences from two kinds of in-plane perturbations within the middle part $(x_\textsc{l},x_\textsc{r})$: 
\begin{align}
   \Hh_\textsc{p}^\textsc{s} &= \sum_{i\in\mathrm{M}} \left[\eta^x_i \sigmax_i + \eta^y_i \sigmay_i \right]\,, \label{Eqn:SpinPerturbation}\\
   \Hh_\textsc{p}^\textsc{f} &= \sum_{i\in\mathrm{M}} \left[\eta^x_i \prod_{j=1}^{i-1} (-\sigmaz_j) \sigmax_i + \eta^y_i \prod_{j=1}^{i-1} (-\sigmaz_j) \sigmay_i \right]\,, \label{Eqn:FermiPerturbation}
\end{align}
where $\eta^{x}_i$ and $\eta^{y}_i$ are perturbation strengths along $x$ and $y$ directions, respectively, both set randomly site by site. Equation~\eqref{Eqn:FermiPerturbation} indeed emulates the conventional local fermionic perturbations from quasi-particle poisoning and breaks Majorana crossings in Fig.~\ref{Fig:Robustness}(b) as expected. On the other hand, in Fig.~\ref{Fig:Robustness}(a) we see that the local spin perturbations [Eq.~\eqref{Eqn:SpinPerturbation}] only shift twofold degeneracy (from external JW Majoranas) away and cannot destroy $\mathbb{Z}_2$ periodicity (even when we extend the random perturbations to the whole spin chain), in stark contrast to topological JJs in superconducting systems.

\section{Conclusions and Outlook}
\label{Sec:Conclusions}

In this paper, we analyzed an Ising-XY-Ising spin link that emulates a topological SNS structure, both analytically and numerically. Our results are summarized in Table~\ref{Tab:Summary} and as follows:

(i) Odd versus even. The iTRS gives rise to the odd-even effect at the lattice level and protects $\mathbb{Z}_4$ ($\mathbb{Z}_2$) fractional spin JE in chains with an odd (even) number of spins, irrespective of ZZ-type interactions. The resulting texture of spin entanglement highlights the effects of the spin current carried by JW Majoranas, whose periodicities can be detected by cQED setup through dispersive readout methods. 

(ii) Lattice versus continuum. By use of the low-energy continuum theory, we analytically solve out the spectra of ABSs and their fermionic wave functions. Nevertheless, the aforementioned odd-even effect can only be observed in a discrete lattice but not in the continuum theory.

(iii) Spin versus fermion. At the lattice level, we identified various symmetries emerging from the spin chain and determine their electronic counterparts, demonstrating that ZZ-type interactions and NN-type interactions affect differently the many-body spectra. One remarkable result is that although $\mathbb{Z}_2$-periodic current can be broken by local fermionic perturbations, spin $\mathbb{Z}_2$ JEs are robust to local spin perturbations.

Our proposal could be implemented in a plethora of spin systems, such as trapped ions \cite{LuN2019}, photonic lattices \cite{RotaPRL2019,RotaPRA2019}, electron spins in quantum dots \cite{ZajacPRA2016}, and magnetic impurities on surfaces \cite{KhajetooriansNRP2019,YangS2019}. In addition, the spin JEs should possibly be simulated and observed in the noisy intermediate-scale quantum computer (e.g. the IBM Q quantum machines) through measuring the correlation functions \cite{PedernalesPRL2014,FrancisPRB2020}. Since $\mathbb{Z}_2$ fractional spin JEs are immune to any local perturbations from arbitrary directions of magnetic field (as long as the chain is still in the topological phase), the ground state can, together with the first excited state, be used to set up a logical qubit: advancing $\phi$ adiabatically by $\pi$ realizes a quantum X gate \cite{TserkovnyakPRA2011,PosskePRL2019}. Alternatively, we can utilize such robustness for quantum memory. In addition, the middle XY chain will be gapped when $|g|>2t$, which prohibits the transport of spin supercurrent. Hence, one may use this feature to engineer a quantum spin transistor based on the JEs \cite{MarchukovNC2016}. 

There are several generalizations of our paper. First, it would be interesting to consider dissipation \cite{PuelPRL2019} (due to, for example, the presence of a magnetic substrate), and evaluate its effects on the various fractional JEs, as well as on the topology of the chain in general. Moreover, the cQED setup proposed here could serve as an engineered environment that can not only monitor the spin flow, but also affect and control it. Second, generalization to multi-junction quantum spin chains, similar to superconducting systems \cite{RiwarNC2016}, which could result in emulating various higher dimensional topological structures. Third, generalization to more complex insulating quantum spin systems, such as 2D quantum (anti)ferromagnets insulators or even quantum spin liquids \cite{ChatterjeePRB2019}, subject to dissipationless spin flows.

\section{Acknowledgments}

We thank Peter Zoller, Dong-Ling Deng, Thore Posske, Tie-Cheng Guo for helpful discussions. This work was supported by the NSFC under the Research Fund for International Young Scientists No.11750110412 and the International Centre for Interfacing Magnetism and Superconductivity with Topological Matter project (M.T.), carried out within the International Research Agendas program of the Foundation for Polish Science co-financed by the European Union under the European Regional Development Fund. S.H. was supported by the Center for Molecular Magnetic Quantum Materials, an Energy Frontier Research Center funded by the U.S. Department of Energy, Office of Science, Basic Energy Sciences under Award No. DE-SC0019330.

\appendix

\section{General Properties of the Spin Chain} \label{Appx:SpinChain}

The generalized 1D anisotropic spin chain Hamiltonian in a transverse field is given by
\begin{align} \label{Appx:HSpinGeneral}
  \Hh_\textsc{g}^\textsc{s}=
  & - J\sum\nolimits_i [ (t_i+\gamma_i) \sigmah^m_i \sigmah^m_{i+1} + (t_i-\gamma_i) \sigmah^{n}_i \sigmah^{n}_{i+1} \nonumber\\
  & + \delta_i \sigmaz_i \sigmaz_{i+1} + g_i \sigmaz_i ]\,.
\end{align}
After the JWT, we obtain the generalized Hamiltonian in the fermionic representation:
\begin{align} \label{Appx:HFermiGeneral}
  \Hh_\textsc{g}^\textsc{f}=
  & - 2J\sum\nolimits_i [ (t_i\ch^\dagger_i \ch_{i+1} + \gamma_i \expm^{-2\im\phi} \ch^\dagger_{i} \ch^\dagger_{i+1} + \mathrm{H.c.}) \nonumber\\
  & + \delta_i (1/2-\ch^\dagger_{i} \ch_{i}-\ch^\dagger_{i+1} \ch_{i+1}+2\nh_{i}\nh_{i+1}) \nonumber\\
  & + g_i( \ch^\dagger_{i} \ch_{i} - 1/2 )]\,,
\end{align}
where the global spin anisotropic angle $\phi$ generating a global gauge transformation $\ch_{i} \rightarrow \ch_{i} \expm^{\im \phi}$. Starting from the non-interacting case $\delta_i=0$, if all the parameters in Eq.~\eqref{Appx:HFermiGeneral} are invariant at every site, we can impose periodic boundary conditions to yield translation symmetry, which does not affect bulk properties. Through applying the Fourier transformation $\ch_k = \sum_j \ch_j \expm^{-\im kaj}/\sqrt{N}$, the Hamiltonian in the momentum space reads:
\begin{align} \label{Appx:HFermiK}
  \Hh_\textsc{g}^{k}=
  & -2J\sum\nolimits_{k} [ (2t\cos ka + g ) \ch^\dagger_{k} \ch_{k} \nonumber\\
  & + \gamma\sin ka(\im\expm^{-2\im\phi} \ch^\dagger_{k} \ch^\dagger_{-k} + \mathrm{H.c.}) -g/2 ]\,,
\end{align}
where $k = 2\pi n/(Na) $ is the wave number with $n$ taking in the range of $( \lfloor -N/2 \rfloor, \lfloor +N/2 \rfloor ]$. Defining a momentum spinor $\hat{C}_k = [ \ch_k,\ch^\dagger_{-k} ]^\textsc{t}$, we write down the BdG Hamiltonian $\Hh_\textsc{g}^{k} = \nicefrac{1}{2} \sum_{k} \hat{C}^\dagger_k \mathcal{H}_\textsc{g}^{k} \hat{C}_k$ with matrix
\begin{equation} \label{Appx:HFermiKBdG}
  \mathcal{H}_\textsc{g}^{k}/2J = -(2t\cos ka + g )\; \rho_z + 2 \gamma \sin ka \;\expm^{-2\im\phi\rho_z} \rho_y\,.
\end{equation}
Now Eq.~\eqref{Appx:HFermiK} can be readily diagonalized into Eq.~\eqref{Eqn:KSpetrum} as $\Hh_\textsc{g}^{k} = \nicefrac{1}{2} \sum\nolimits_{k} \hat{D}^\dagger_k \epsilon_k \rho_z \hat{D}_k = \sum\nolimits_k \epsilon_k (\dch^\dagger_k\dch_k - \nicefrac{1}{2}),$ by introducing the Bogoliubov quasi-particle $\hat{D}_k = [ \dch_k,\dch^\dagger_{-k} ]^\textsc{t}$ as $\dch_k = \expm^{+\im \phi} \sin (\theta_k/2) \; \ch_k - \im\,\expm^{-\im \phi} \cos(\theta_k/2) \; \ch^\dagger_{-k}$ with $\theta_k = \arctan [ 2\gamma\sin ka/(2t\cos{ka}+g) ].$
We can use $\theta_k$ to define the topological invariant by the winding number
\begin{equation}
  \mathcal{W}=\frac{1}{2\pi}\oint\dm\theta_k=\frac{1}{2\pi}\int_{\mathrm{BZ}} \frac{\dm\theta_k}{\dm k} \dm k = \Theta (2t - |g|)\,,
\end{equation}
where $\Theta$ is the Heaviside step function. When $g<|2t|$ the bulk is in the topological phase with $\mathcal{W}=1$, which means if the chain was cut at a point, two unpaired Majorana modes would appear at the ends of it. However, if $\mathcal{W}=0$ the bulk will lie in the trivial phase and the edge modes disappear, which is known as the bulk-edge correspondence.

When the spin chain consists of different parametric parts, $k$ is not a good quantum number anymore, we should come back to the real space. Especially for the non-interacting case $\delta_i=0$, Eq.~\eqref{Appx:HFermiGeneral} is reduced into the single-particle form $\mathcal{H}_\textsc{g}^\textsc{f}$ shown in Eq.~\eqref{Eqn:HBdGFermiGeneral}. By use of the PHS as $\{\mathcal{H}_\textsc{g}^\textsc{f}, \mathcal{C}_\textsc{f}\}=0$, $\mathcal{C}_\textsc{f}=\rho_x \mathcal{K}$, for every eigenvector $\Phi^+_n = [u_n(1),\dots,u_n(\mathrm{N}),v_n(1),\dots,v_n(\mathrm{N})]^\textsc{t}$ with positive energy $+\epsilon_n$, there is a corresponding eigenvector $\Phi^-_n = \mathcal{C}_\textsc{f}\Phi^+_n = [v^*_n(1),\dots,v^*_n(\mathrm{N}),u^*_n(1),\dots,u^*_n(\mathrm{N})]^\textsc{t}$ for the negative energy $-\epsilon_n$. Therefore, $\mathcal{H}_\textsc{g}^\textsc{f}$ can be diagonalized as $\Hh_\textsc{g}^\textsc{f} = \nicefrac{1}{2}\hat{C}^\dagger \mathcal{H}_\textsc{g}^\textsc{f} \hat{C} = \nicefrac{1}{2}\hat{C}^\dagger \mathcal{P} \mathcal{E} \mathcal{P}^\dagger \hat{C} = \nicefrac{1}{2} \hat{D}^\dagger \mathcal{E} \hat{D} = \sum_n \epsilon_n ( \dch^\dagger_n \dch_n - \nicefrac{1}{2} )$ by the Bogoliubov quasi-particle $\hat{D} = (\dch_1, \dch_2, \dots, \dch_\textsc{n}, \dch^\dagger_1, \dch^\dagger_2, \dots, \dch^\dagger_\textsc{n})^\textsc{t}$, where $\mathcal{E} = \sum_n \rho_z \otimes \epsilon_n\ket{n}\bra{n}$, and $\mathcal{P} \equiv [\Phi^+_1,\dots,\Phi^+_\textsc{n},\Phi^-_1,\dots,\Phi^-_\textsc{n}]$ is constructed by their corresponding eigenvectors, whose column vectors and row vectors should be orthonormal:
\begin{align} \label{Appx:Orth}
  \sum\nolimits_i \left[ u^*_m(i) u_n(i) + v^*_m(i) v_n(i) \right] &= \delta_{m,n}\,, \nonumber\\
  \sum\nolimits_n \left[ u^*_n(i) u_n(j) + v_n(i) v^*_n(j) \right] &= \delta_{i,j}\,.
\end{align}
Since $\hat{D}=\mathcal{P}^\dagger\hat{C}, \hat{C}=\mathcal{P}\hat{D}$, the transformation between quasi-particles and fermions is given by
\begin{align} \label{Appx:CtoD}
  \dch_n &= \sum\nolimits_i [ u^*_n(i) \ch_i + v^*_n(i) \ch^\dagger_i ], \nonumber\\
  \ch_i  &= \sum\nolimits_n [ u_n(i) \dch_n + v^*_n(i) \dch^\dagger_n ]\,.
\end{align}
If there are interacting terms $\delta_i\neq0$ in Eq.~\eqref{Appx:HFermiGeneral}, above single-particle method fails since the Hamiltonian will not be quadratic anymore. Under this circumstance, we have to stay in the spin space and apply brute-force diagonalization on a $2^N\times 2^N$ matrix of Eq.~\eqref{Appx:HSpinGeneral} to solve out the many-body spectrum directly.

\section{Low-Energy Continuum Theory}
\label{Appx:Continuum}

\subsection{wave functions Near the Critical Point} 
\label{Appx:CriticalWFs}

We can diagonalize the low-energy continuous Eq.~\eqref{Eqn:HCriticalGeneral} as $\mathcal{H}_\textsc{g}^\textsc{c}\Phi(x) = \epsilon\Phi(x)$ by solving out differential equations of the two-component wave function $\Phi(x) = \left[u(x),v(x)\right]^\textsc{t}$, whose generalized expressions are shown as
\begin{align} \label{Appx:GeneralWFs}
  u(x) =
  & \expm^{-\im\phi} (+\mathrm{C}_1 \cos\mathrm{U}\, \expm^{+\mathrm{K}^\textsc{+} x} + \mathrm{C}_2 \sin\mathrm{V}\, \expm^{+\mathrm{K}^\textsc{-} x} \nonumber\\
  &+ \mathrm{C}_3 \cos\mathrm{U}\, \expm^{-\mathrm{K}^\textsc{+} x} + \mathrm{C}_4 \sin\mathrm{V}\, \expm^{-\mathrm{K}^\textsc{-} x} )\,, \nonumber\\
  v(x) =
  & \expm^{+\im\phi} (-\mathrm{C}_1 \cos\mathrm{V}\, \expm^{+\mathrm{K}^\textsc{+} x} - \mathrm{C}_2 \sin\mathrm{U}\, \expm^{+\mathrm{K}^\textsc{-} x} \nonumber\\
  &+ \mathrm{C}_3 \cos\mathrm{V}\, \expm^{-\mathrm{K}^\textsc{+} x} + \mathrm{C}_4 \sin\mathrm{U}\, \expm^{-\mathrm{K}^\textsc{-} x} )\,,
\end{align}
where $\mathrm{K}^\ssub{\pm}=\sqrt{\Gamma-\Omega\pm\Lambda}/ta$, $\mathrm{U}=\arccos[(\Lambda-\Xi)/\Gamma]/2$, $\mathrm{V}=\arccos[(\Lambda+\Xi)/\Gamma]/2$, $\Lambda=\sqrt{\Gamma^2+\Xi^2-2\Gamma\Omega}$, $\Gamma=2\gamma^{2}$, $\Omega=t(2t+g)$, $\Xi=t\epsilon/2J$ are introduced for simplicity. Additionally, $\mathrm{K}^\textsc{+}=2\gamma\cos\mathrm{U}\cos\mathrm{V}/ta$, $\mathrm{K}^\textsc{-}=2\gamma\sin\mathrm{U}\sin\mathrm{V}/ta$. Applying infinite boundary conditions on Eqs.~\eqref{Appx:GeneralWFs}, the right part wave functions are defined by setting $\mathrm{C}_1=\mathrm{C}_2=0$, and the left part of the wave functions are obtained by setting $\mathrm{C}_3=\mathrm{C}_4=\phi=0$. The middle part is a special case of $\phi=\gamma=0$; one could reduce $\mathrm{K}^\ssub{\pm} \rightarrow \im \sqrt{\Omega\mp\Xi}/ta \equiv \im \mathrm{K}_\textsc{m}^\ssub{\mp}$ and find $\mathrm{K}_\textsc{m}^\textsc{+}=2\gamma\sin\mathrm{U}\cos\mathrm{V}/ta$, $\mathrm{K}_\textsc{m}^\textsc{-}=2\gamma\cos\mathrm{U}\sin\mathrm{V}/ta$ after taking the limit $\gamma\rightarrow0$. We are only interested in the ABSs, whose eigenvalues lie within the gap, i.e. $|\epsilon|<2J(2t+g)\Leftrightarrow|\Xi|<\Omega$, which ensures $\mathrm{K}_\textsc{m}^\ssub{\pm}$ to be real. By introducing a new set of coefficients $\mathrm{C}_5,\mathrm{C}_6,\mathrm{C}_7,\mathrm{C}_8$ in the middle region, the explicit wave functions are shown as
\begin{align} \label{Appx:MiddleWFs}
  u_\textsc{m}(x) 
  = 1/\sqrt{\mathrm{K}_\textsc{m}^\textsc{+}}\times(\mathrm{C}_5 \expm^{+\im\mathrm{K}_\textsc{m}^\textsc{+} x} + \mathrm{C}_6 \expm^{-\im\mathrm{K}_\textsc{m}^\textsc{+} x})\,, \nonumber\\
  v_\textsc{m}(x) 
  = 1/\sqrt{\mathrm{K}_\textsc{m}^\textsc{-}}\times(\mathrm{C}_7 \expm^{+\im\mathrm{K}_\textsc{m}^\textsc{-} x} + \mathrm{C}_8 \expm^{-\im\mathrm{K}_\textsc{m}^\textsc{-} x})\,.
\end{align}
The above wave functions have been normalized by the square root of wave numbers to maintain the quasi-particle current \cite{BeenakkerPRL1991}. Through imposing continuity and current conservation conditions at two interfaces presented in Appendix~\ref{Appx:WFBC}, we obtain the left- and right-scattering matrices $\mathcal{S}_\textsc{l}^\textsc{c}=\mathcal{S}(-1,0), \mathcal{S}_\textsc{r}^\textsc{c}=\mathcal{S}(+1,\phi)$ with
\begin{equation}
\mathcal{S}(\tau,\phi)=\frac{1}{\mathcal{S}^*_0}
\begin{bmatrix}
\mathcal{S}_1 & \im\tau\expm^{-2\im\phi}\mathcal{S}_2 \\ 
\im\tau\expm^{+2\im\phi}\mathcal{S}_2 & \mathcal{S}^*_1 \end{bmatrix}\,,
\end{equation}
and the entries are defined as
\begin{align*}
\mathcal{S}_0&=\sin\beta(1+\zeta^2-2\zeta\expm^{\im\beta})-2\im\expm^{\im\beta} (\sin^2\alpha-\sin^2\beta)\,,\\
\mathcal{S}_1&=-\sin\beta[1+\zeta^2-2\zeta(\cos\beta+\im \sin\alpha)]\,,\\
\mathcal{S}_2&=2\sin\alpha\sqrt{\sin^2\alpha-\sin^2\beta}\,,
\end{align*}
where $\alpha=\mathrm{U}+\mathrm{V}$, $\beta=\mathrm{U}-\mathrm{V}$, $\zeta=\lambda/\gamma$. The waves at the two interfaces only contain different factors caused by the middle wave number $\mathrm{K}_\textsc{m}^\ssub{\pm}=\sqrt{\Omega\pm\Xi}/ta$, which is described by scattering matrix $\mathcal{S}_\textsc{m}^\textsc{c}=\exp(\im{\rho_z} \mathrm{K}_\textsc{m}^{\rho_z} L)$. Notice that such a wave function factor will be canceled out due to Andreev reflection after traveling for one loop, which enforces $\mathrm{det} (\mathbbm{1}-\mathcal{S}_\textsc{m}^\textsc{c} \mathcal{S}_\textsc{r}^\textsc{c} \mathcal{S}_\textsc{m}^\textsc{c} \mathcal{S}_\textsc{l}^\textsc{c})=0$ and gives the energy transcendental Eq.~\eqref{Eqn:CriticalDET} for the ABSs. The wave-function coefficients are then determined by normalization condition $\int |u_n(x)|^2 + |v_n(x)|^2 \dm x = 1$, and the Hamiltonian is diagonalized into $\sum_n \epsilon_n ( \dch^\dagger_n \dch_n - \nicefrac{1}{2} )$ by Bogoliubon $\dch_n$, whose transformation with field operator is given by
\begin{equation}
  \dch_n = \int \dm x\; \Phi^\dagger_n(x)\Ph(x),\quad
  \Ph(x)=\sum\nolimits_{n} \Phi_n(x)\dch_n.
\end{equation}
Recall Eq.~\eqref{Eqn:HCriticalGeneral} holds the PHS as $\{\mathcal{H}_\textsc{g}^\textsc{c}, \mathcal{C}_\textsc{c}\}=0$ by the operator $\mathcal{C}_\textsc{c} = \rho_x \mathcal{K}$, thus $\mathcal{C}_\textsc{c} \Phi_n(x) = \left[v^*_n(x), u^*_n(x)\right]^\textsc{t} \equiv \Phi_{-n}(x)$ is the wave function for $-\epsilon_n\equiv\epsilon_{-n}$. It is worthwhile to point out that it is the branch cut of $\mathrm{V}$ on the Riemann surface that takes great effect on the quantum phase transition, i.e., $\mathrm{V}\rightarrow-\arccos[(\Lambda+\Xi)/\Gamma]/2$ with an additional minus sign across the critical point, which prohibits the zero-mode solution of Majoranas.

\subsection{wave functions in the Deep Topological Regime}
\label{Appx:DeepWFs}

Owing to $[\mathcal{H}_\textsc{g}^\textsc{d}, \tau_z]=0$, it is more convenient for us to decompose the Hilbert space in two $\tau_z$ eigen-sectors $\tau=\pm 1$ and solve out Eq.~\eqref{Eqn:HDeepGeneral} as $\mathcal{H}_\textsc{g}^\textsc{d} \Phi^\tau(x) = \epsilon^\tau \Phi^\tau(x)$  with their corresponding eigenfunctions $\Phi^\textsc{+}(x)=[u^\textsc{+}(x), 0, v^\textsc{+}(x), 0]^\textsc{t}, \Phi^\textsc{-}(x)=[0, u^\textsc{-}(x), 0, v^\textsc{-}(x)]^\textsc{t}$, whose explicit expressions are shown as
\begin{align}
	u^\tau(x) &= \expm^{-\im \phi} (\mathrm{C}_1\, \expm^{-\im\mathrm{W}} \expm^{+\tau \mathrm{K} x} + \mathrm{C}_2\, \expm^{+\im\mathrm{W}} \expm^{-\tau\mathrm{K} x})\,, \nonumber\\
	v^\tau(x) &= \expm^{+\im \phi} (\mathrm{C}_1\, \expm^{+\im\mathrm{W}} \expm^{+\tau \mathrm{K} x} + \mathrm{C}_2\, \expm^{-\im\mathrm{W}} \expm^{-\tau\mathrm{K} x})\,,
\end{align}
where $\mathrm{K}=\sqrt{\Delta^2-\mathrm{E}^2}/\Upsilon$, $\mathrm{W}=\arccos(\mathrm{E}/\Delta)/2$, $\mathrm{E}=\epsilon/2J$ are introduced for simplicity. The wave functions of left and right parts only contain the exponential decaying branches due to infinite boundary conditions, while the middle part is the case of $\phi=\gamma=0$ where $\mathrm{K}=\im\mathrm{E}/\Upsilon \equiv \im \mathrm{K}_\textsc{m}$. Since $\epsilon_\ssub{gap} \rightarrow 4J\gamma, \Delta\rightarrow2\gamma$ in the deep topological regime, $|\mathrm{E}|<\Delta$ will be always valid for the ABSs. The explicit middle wave functions are shown as $u_\textsc{m}^\tau(x) = \mathrm{C}_3 \exp \left( +\im\tau \mathrm{K}_\textsc{m} x \right), v_\textsc{m}^\tau(x) = \mathrm{C}_4 \exp \left( -\im\tau \mathrm{K}_\textsc{m} x \right)$ with two new coefficients. Applying continuity conditions at two interfaces $x_\textsc{l,r}$ on the wave functions of each eigen-sector, we can obtain the left- and right-scattering matrices $\mathcal{S}_\textsc{l}^\textsc{d}=\exp (-2\im\mathrm{W}) \rho_x$, $\mathcal{S}_\textsc{r}^\textsc{d}=\exp (-2\im\mathrm{W}) \expm^{-2\im\phi\rho_z} \rho_x$, where we denote $\mathrm{W}=\arccos(\mathrm{E}/\Delta)/2$, $\mathrm{E}=\epsilon/2J$. The scattering matrix of the middle part is only determined by the middle wave number $\mathrm{K}_\textsc{m}=\mathrm{E}/\Upsilon$ as $\mathcal{S}_\textsc{m}^\textsc{d}=\exp (\im \mathrm{K}_\textsc{m} L) \expm^{\im k_\textsc{f}\rho_z}$. The solvability equation $\mathrm{det} (\mathbbm{1}-\mathcal{S}_\textsc{m}^\textsc{d} \mathcal{S}_\textsc{r}^\textsc{d} \mathcal{S}_\textsc{m}^\textsc{d} \mathcal{S}_\textsc{l}^\textsc{d})=0$ of the Andreev reflection gives the energy transcendental Eq.~\eqref{Eqn:DeepDET} for the ABSs. By use of the normalization condition, the full normalized wave functions for the whole chain are expressed as
\begin{align}
	u_n^\tau(x) &= \mathcal{A}_n\, (-1)^n \; \expm^{ -\mathrm{K} |x-l(x)| } \expm^{ +\im \tau \mathrm{K}_\textsc{m} l(x) }\,, \nonumber\\
	v_n^\tau(x) &= \mathcal{A}_n\, \expm^{+\im \phi}\! \times \expm^{ -\mathrm{K} |x-l(x)| } \expm^{ -\im \tau \mathrm{K}_\textsc{m} l(x) }\,,
\end{align}
where $\mathcal{A}_n=1/\sqrt{2(L+1/\mathrm{K})}$ is the normalization factor, $l(x)=x$ for $x\leq|L/2|$ and $\mathrm{sgn}(x)L/2$ for $x>|L/2|$. Using the PHS as $\{\mathcal{H}_\textsc{g}^\textsc{d}, \mathcal{C}_\textsc{d}\}=0, \mathcal{C}_\textsc{d}=\rho_y\tau_y\mathcal{K}$, Eq.~\eqref{Eqn:HDeepGeneral} is diagonalized into $\sum_{n,\tau} \epsilon_n^{\tau} (\dch^{\tau\dagger}_n \dch^\tau_n-\nicefrac{1}{2})$ with:
\begin{equation}
\dch^\tau_n=\int \dm x\; \Phi^{\tau\dagger}_n(x) \Ph(x)\,, \quad \Ph(x)=\sum_{n,\tau} \Phi^\tau_n(x) \dch^\tau_n\,.
\end{equation}

\subsection{Boundary Conditions Near the Critical Point} \label{Appx:WFBC}

We can add a fictitious barrier potential $\lambda a \delta(x-x_\ssub{\pm})$ into Eq.~\eqref{Eqn:HCriticalGeneral} to emulate the imperfect connections between different parts (we denote $+, -$ for R, L respectively to generalize the expressions of two junction sites in the following statements). Around two interfaces, the stationary Schr\"{o}dinger equation requires:
\begin{align*}
  \mathcal{H}_\textsc{g}^\textsc{c} \Phi(x) =& -2J \{ [ 2t+g+\lambda a \delta(x-x_\ssub{\pm})+ta^2 \partial^2_x ]\rho_z\\
  &+ \im \gamma a [ \Theta(\pm x \mp x_\ssub{\pm}),\; \partial_x ]_\textsc{+} \rho_y \} \Phi(x) = \epsilon \Phi(x)\,,
\end{align*}
where the phase $\phi$ is absorbed in $\gamma$ temperately, the anti-commutator parentheses $[ \Theta(\pm x \mp x_\ssub{\pm}),\; \partial_x ]_\textsc{+}$ can be calculated into $2\Theta(\pm x \mp x_\ssub{\pm}) \partial_x \pm \delta(x-x_\ssub{\pm})$. Moving the second-order derivative term to the left-hand side and integrating the whole equation around the junction sites by an infinitesimal parameter, we find
\begin{equation}
  ta\begin{bmatrix}
	+ u'_\ssub{\pm}(x_\ssub{\pm}) - u'_\textsc{m}(x_\ssub{\pm})\\
	- v'_\ssub{\pm}(x_\ssub{\pm}) + v'_\textsc{m}(x_\ssub{\pm})
	\end{bmatrix}
  =\begin{bmatrix}
	\mp \lambda u_\textsc{m}(x_\ssub{\pm}) - \gamma u_\ssub{\pm}(x_\ssub{\pm})\\
	\pm \lambda v_\textsc{m}(x_\ssub{\pm}) + \gamma v_\ssub{\pm}(x_\ssub{\pm})
	\end{bmatrix}.
\end{equation}
Replacing subscript $+, -$ back into R, L and specifying the value of $\gamma, \phi$ in different parts (releasing $\phi$ from $\gamma$), we obtain the current conservation conditions:
\begin{align} \label{Appx:BCsGeneral}
  ta\,u'_\textsc{m}(x_\textsc{l}) + \lambda u_\textsc{m}(x_\textsc{l}) &= ta\,u'_\textsc{l}(x_\textsc{l}) + \gamma v_\textsc{l}(x_\textsc{l}) \,, \nonumber\\
  ta\,v'_\textsc{m}(x_\textsc{l}) + \lambda v_\textsc{m}(x_\textsc{l}) &= ta\,v'_\textsc{l}(x_\textsc{l}) + \gamma u_\textsc{l}(x_\textsc{l}) \,, \nonumber\\
  ta\,u'_\textsc{m}(x_\textsc{r}) - \lambda u_\textsc{m}(x_\textsc{r}) &= ta\,u'_\textsc{r}(x_\textsc{r}) + \gamma \expm^{-2\im\phi}v_\textsc{r}(x_\textsc{r}) \,, \nonumber\\
  ta\,v'_\textsc{m}(x_\textsc{r}) - \lambda v_\textsc{m}(x_\textsc{r}) &= ta\,v'_\textsc{r}(x_\textsc{r}) + \gamma \expm^{+2\im\phi}u_\textsc{r}(x_\textsc{r}) \,,
\end{align}
together with four trivial wave function continuity conditions $u_\textsc{l}(x_\textsc{l}) = u_\textsc{m}(x_\textsc{l})$, $v_\textsc{l}(x_\textsc{l}) = v_\textsc{m}(x_\textsc{l})$, $u_\textsc{r}(x_\textsc{r}) = u_\textsc{m}(x_\textsc{r})$, $v_\textsc{r}(x_\textsc{r}) = v_\textsc{m}(x_\textsc{r})$. When $\lambda=0$, Eqs.~\eqref{Appx:BCsGeneral} impose perfect coupling boundary conditions while if $\lambda\rightarrow\infty$ the three parts in our chain system are independent, and the $\phi$ dependence will be suppressed. One could use $\lambda \sim (t-\mathbbm{t})/\mathbbm{t}$ as a fitting function for the mapping between the lattice and the continuum model, while the explicit formula is varied with different parameter ranges, which is beyond the scope of this paper.

\section{Spin Correlation Functions} \label{Appx:CoFunc}

By use of transformation Eqs.~\eqref{Appx:CtoD} and orthonormality conditions of wave functions Eqs.~\eqref{Appx:Orth}, we define two operators,
\begin{align}
  \hat{A}_i &= \ch^\dagger_i + \ch_i = \sum\nolimits_n [ a^*_n(i) \dch^\dagger_n + a_n(i) \dch_n ]\,, \nonumber\\
  \hat{B}_i &= \ch^\dagger_i - \ch_i = \sum\nolimits_n [ b^*_n(i) \dch^\dagger_n - b_n(i) \dch_n ]\,,
\end{align}
with $a_n(i)=u_n(i) + v_n(i)$, $b_n(i)=u_n(i) - v_n(i)$, and their the expectation values by pairs $M_{i,j}\equiv\braket{\hat{A}_i \hat{A}_j}$, $N_{i,j}\equiv\braket{\hat{B}_i \hat{B}_j}$, $ G_{i,j}\equiv\braket{\hat{B}_i \hat{A}_j}$ are calculated as 
\begin{align*}
  M_{i,j}&= +\delta_{ij} + 2\im\, \mathrm{Im} \sum\nolimits_n [ u_n(i)a^*_n(j) + a^*_n(i) a_n(j) f_n ]\,,\\
  N_{i,j}&= -\delta_{ij} - 2\im\, \mathrm{Im} \sum\nolimits_n [ u_n(i)b^*_n(j) + b^*_n(i) b_n(j) f_n ]\,,\\
  G_{i,j}&= +\delta_{ij} - 2\,\;  \mathrm{Re} \sum\nolimits_n [ u_n(i)a^*_n(j) - b^*_n(i) a_n(j) f_n ]\,,
\end{align*}
where $f_n\equiv\braket{\dch^\dagger_n \dch_n}$ is the occupation number of quasi-particles. These expressions are different from Refs.~\cite{LiebAP1961,OsbornePRA2002,Sachdev2011} as a result of the imaginary parts of the wave functions stemming from the spin supercurrent in Eq.~\eqref{Eqn:SpinCurrent}:
\begin{equation}\label{Appx:SpinCurrent}
  \braket{\hat{J_z}}/(-2Jt) = \mathrm{Im} [N_{i,i+1}-M_{i,i+1}] \ . \
\end{equation}
It is easy to find $\braket{\hat{B}_i \hat{A}_j}=-\braket{\hat{A}_j \hat{B}_i}$,  $\braket{\hat{A}_i \hat{A}_j}=\braket{\hat{A}_j \hat{A}_i}^*$, $\braket{\hat{B}_i \hat{B}_j}=\braket{\hat{B}_j \hat{B}_i}^*$ and obtain $\braket{\sigmaz_i} = \braket{\hat{B}_i \hat{A}_i} = G_{i,i}\,$,
$\braket{\sigmaz_i \sigmaz_j} = \braket{\hat{B}_i \hat{A}_i \hat{B}_j \hat{A}_j} = G_{i,i}G_{j,j} - G_{i,j}G_{j,i} - N_{i,j}M_{i,j}\,$.
However, it is not so straightforward to obtain the following correlators at arbitrary length $k=|i-j|$:
\begin{align}\label{Appx:TwoAB}
\braket{\sigmax_i\sigmax_j} &= \;+ \braket{\hat{B}_{i} \hat{A}_{i+1} \hat{B}_{i+1} \cdots \hat{A}_{j-1} \hat{B}_{j-1} \hat{A}_{j}}\,, \nonumber\\
\braket{\sigmay_i\sigmay_j} &= \;- \braket{\hat{A}_{i} \hat{A}_{i+1} \hat{B}_{i+1} \cdots \hat{A}_{j-1} \hat{B}_{j-1} \hat{B}_{j}}\,, \nonumber\\
\braket{\sigmax_i\sigmay_j} &= -\im\braket{\hat{B}_{i} \hat{A}_{i+1} \hat{B}_{i+1} \cdots \hat{A}_{j-1} \hat{B}_{j-1} \hat{B}_{j}}\,, \nonumber\\
\braket{\sigmay_i\sigmax_j} &= -\im\braket{\hat{A}_{i} \hat{A}_{i+1} \hat{B}_{i+1} \cdots \hat{A}_{j-1} \hat{B}_{j-1} \hat{A}_{j}}\,,
\end{align}
which will be expanded into $(2k-1){!}{!}$ terms according to Wick theorem. Those correlators are found to be systematically expressed as the Pfaffian
\begin{align}
\braket{\sigmax_i\sigmax_j} &= +\,(-1)^{k(k-1)/2} \;\mathrm{pf}(\mathcal{Q}^{xx}_{ij}) \,, \nonumber\\
\braket{\sigmay_i\sigmay_j} &= +\,(-1)^{k(k-1)/2} \;\mathrm{pf}(\mathcal{Q}^{yy}_{ij}) \,, \nonumber\\
\braket{\sigmax_i\sigmay_j} &= -\im(-1)^{k(k-1)/2} \;\mathrm{pf}(\mathcal{Q}^{xy}_{ij})\,, \nonumber\\
\braket{\sigmay_i\sigmax_j} &= +\im(-1)^{k(k-1)/2} \;\mathrm{pf}(\mathcal{Q}^{yx}_{ij})\,,
\end{align}
of the following well-organized $2k\times 2k$ skew-symmetric matrices \cite{CaianielloNC1952,BarouchPRA1971}:

\begin{equation*}
\mathcal{Q}^{xx}_{ij}=
\begin{bmatrix}
\mathcal{N}^{xx}_{ij} & \mathcal{G}^{xx}_{ij} \\
-{\mathcal{G}^{xx}_{ij}}^\textsc{t} & \mathcal{M}^{xx}_{ij} \\
\end{bmatrix}\,,\quad
\mathcal{Q}^{yy}_{ij}=
\begin{bmatrix}
\mathcal{M}^{yy}_{ij} & \mathcal{G}^{yy}_{ij} \\
-{\mathcal{G}^{yy}_{ij}}^\textsc{t} & \mathcal{N}^{yy}_{ij} \\
\end{bmatrix}\,,
\end{equation*}
with their corresponding blocks

\begin{widetext}

\begin{equation*}
\mathcal{G}^{xx}_{ij}=
\begin{bmatrix}
G_{i,i+1}& \cdots & G_{i,j-1} & G_{i,j}\\
G_{i+1,i+1}& \cdots & G_{i+1,j-1} & G_{i+1,j}\\
\vdots & \ddots & \vdots & \vdots \\
G_{j-1,i+1}& \cdots & G_{j-1,j-1} & G_{j-1,j}
\end{bmatrix}\,, \quad
\mathcal{G}^{yy}_{ij}=
\begin{bmatrix}
G_{i+1,i}& \cdots & G_{j-1,i} & G_{j,i}\\
G_{i+1,i+1}& \cdots & G_{j-1,i+1} & G_{j,i+1}\\
\vdots & \ddots & \vdots & \vdots \\
G_{i+1,j-1}& \cdots & G_{j-1,i-1} & G_{j,j-1}
\end{bmatrix}\,,
\end{equation*}

\begin{equation*}
\mathcal{M}^{xx}_{ij}=
\begin{bmatrix}
0 & M_{i+1,i+2}& \cdots  & M_{i+1,j}\\
-M_{i+1,i+2}& 0 & \ddots  &  \vdots\\
\vdots & \ddots & 0 & M_{j-1,j} \\
-M_{i+1,j} & \cdots & M_{j-1,j}  & 0
\end{bmatrix}\,, \quad
\mathcal{M}^{yy}_{ij}=
\begin{bmatrix}
0 & M_{i,i+1}& \cdots  & M_{i,j-1}\\
-M_{i,i+1}& 0 & \ddots  &  \vdots\\
\vdots & \ddots & 0 & M_{j-2,j-1} \\
-M_{i,j-1} & \cdots & M_{j-2,j-1}  & 0
\end{bmatrix}\,, 
\end{equation*}

\begin{equation*}
\mathcal{N}^{xx}_{ij}=
\begin{bmatrix}
0 & N_{i,i+1}& \cdots  & N_{i,j-1}\\
-N_{i,i+1}& 0 & \ddots  &  \vdots\\
\vdots & \ddots & 0 & N_{j-2,j-1} \\
-N_{i,j-1} & \cdots & N_{j-2,j-1}  & 0
\end{bmatrix}\,, \quad
\mathcal{N}^{yy}_{ij}=
\begin{bmatrix}
0 & N_{i+1,i+2}& \cdots  & N_{i+1,j}\\
-N_{i+1,i+2}& 0 & \ddots  &  \vdots\\
\vdots & \ddots & 0 & N_{j-1,j} \\
-N_{i+1,j} & \cdots & N_{j-1,j}  & 0
\end{bmatrix}\,.
\end{equation*}

Through observing Eq.~\eqref{Appx:TwoAB}, the correlators of $\braket{\sigmax_i\sigmay_j}$, $\braket{\sigmay_i\sigmax_j}$ only differ on the last operator from $\braket{\sigmax_i\sigmax_j}$, $\braket{\sigmay_i\sigmay_j}$, hence we can calculate $\mathcal{Q}^{xy}_{ij}$, $\mathcal{Q}^{yx}_{ij}$ by replacing the last column and its corresponding transpose row $\boxed{\cdots}$ of $\mathcal{Q}^{xx}_{ij}$, $\mathcal{Q}^{yy}_{ij}$, respectively:

\begin{equation*}
\mathcal{Q}^{xx}_{ij}=
\begin{bmatrix}
& \cdots & G_{i,j}\\
& \cdots & G_{i+1,j}\\
& \cdots & \vdots \\
& \cdots & G_{j-1,j}\\
& \cdots & M_{i+1,j}\\
& \cdots & \vdots\\
& \cdots & M_{j-1,j} \\
& \boxed{GM} & 0
\end{bmatrix}
\Rightarrow
\begin{bmatrix}
& \cdots & N_{i,j}\\
& \cdots & N_{i+1,j}\\
& \cdots & \vdots \\
& \cdots & N_{j-1,j}\\
& \cdots & -G_{j,i+1}\\
& \cdots & \vdots\\
& \cdots & -G_{j,j-1} \\
& \boxed{NG} & 0
\end{bmatrix}
\equiv\mathcal{Q}^{xy}_{ij}\,; \quad
\mathcal{Q}^{yy}_{ij}=
\begin{bmatrix}
& \cdots & G_{j,i}\\
& \cdots & G_{j,i+1}\\
& \cdots & \vdots \\
& \cdots & G_{j,j-1}\\
& \cdots & N_{i+1,j}\\
& \cdots & \vdots\\
& \cdots & N_{j-1,j} \\
& \boxed{GN} & 0
\end{bmatrix}
\Rightarrow
\begin{bmatrix}
& \cdots & M_{j,i}\\
& \cdots & M_{j,i+1}\\
& \cdots & \vdots \\
& \cdots & M_{j,j-1}\\
& \cdots & G_{i+1,j}\\
& \cdots & \vdots\\
& \cdots & G_{j-1,j} \\
& \boxed{MG} & 0
\end{bmatrix}
\equiv\mathcal{Q}^{yx}_{ij}\,.
\end{equation*}

When the twisting angle is zero, the spin supercurrent vanishes with $\braket{\sigmax_i\sigmay_j}=\braket{\sigmay_i\sigmax_j}=0$. Furthermore, block-diagonal terms in $\mathcal{Q}^{xx}_{ij}$, $\mathcal{Q}^{yy}_{ij}$ are also found out to be zero. In this special case, $\braket{\sigmax_i\sigmax_j}$ and $\braket{\sigmay_i\sigmay_j}$ are reduced into $\mathrm{det} (\mathcal{G}^{xx}_{ij})$ and $\mathrm{det} (\mathcal{G}^{yy}_{ij})$, respectively, which agree with previous formulas used in Refs.~\cite{LiebAP1961,OsbornePRA2002,Sachdev2011}.

\end{widetext}

\bibliography{references}

\end{document}